%% file: paper.tex
\documentclass[sigconf]{acmart}
\AtBeginDocument{
  }

\copyrightyear{2026}
\acmYear{2026}
\setcopyright{cc}
\setcctype{by}
\acmConference[CCS '26]{Proceedings of the 2026 ACM SIGSAC Conference on Computer and Communications Security}{November 15--19, 2026}{The Hague, Netherlands}
\acmBooktitle{Proceedings of the 2026 ACM SIGSAC Conference on Computer and Communications Security (CCS '26), November 15--19, 2026, The Hague, Netherlands}
\acmDOI{10.1145/3830454.3846582}
\acmISBN{979-8-4007-2871-6/2026/11}

\input{preamble}
\graphicspath{{figs/}}

\begin{document}

\title[Where Have All the Firewalls Gone? Security Consequences of Residential
IPv6 Transition]{Where Have All the Firewalls Gone?\\Security Consequences of Residential IPv6 Transition}

\author{Erik Rye}
\affiliation{
  \institution{Johns Hopkins University}
  \city{}
  \state{}
  \country{}
}
\email{rye@jhu.edu}

\author{Dave Levin}
\affiliation{
  \institution{University of Maryland}
  \city{}
  \state{}
  \country{}
}
\email{dml@cs.umd.edu}

\author{Robert Beverly}
\affiliation{
  \institution{San Diego State University}
  \city{}
  \state{}
  \country{}
}
\email{rbeverly@sdsu.edu}

\begin{abstract}
    \input{dml-abstract}
\end{abstract}



\maketitle

\input{intro}
\input{background}
\input{methodology}
\input{results}
\input{casestudies}
\input{mitigations}
\input{conclusion}

\section*{Acknowledgments}
We thank Chris Eagle for early feedback.
This research was partially
supported by the NSF under grants CNS-2323193 and OAC-2613546.
Views and conclusions contained in this document are those of the
authors and should not be interpreted as representing the
official policies, either expressed or implied, of NSF or the
U.S.\ government.

\bibliographystyle{ACM-Reference-Format}
\bibliography{conferences,refs}

\appendix
\input{appendix}

\begin{acronym}
  \acro{AS}{Autonomous System}
  \acrodefplural{AS}[ASes]{Autonomous Systems}
  \acro{ASN}{\ac{AS} Number}
  \acro{AUP}{Acceptable Use Policy}
  \acro{BGP}{Border Gateway Protocol}
  \acro{CCPA}{California Consumer Privacy Act}
  \acro{CDN}{Content Distribution Network}
  \acro{CPE}{Customer Premises Equipment}
  \acro{CSIRT}{Computer Security Incident Response Team}
  \acro{DAD}{Duplicate Address Detection}
  \acro{EUI}{Extended Unique Identifier}
  \acro{GDPR}{General Data Protection Regulation}
  \acro{IoT}{Internet of Things}
  \acro{ISP}{Internet Service Provider}
  \acro{IID}{Interface Identifier}
  \acro{LAN}{Local Area Network}
  \acro{NIC}{Network Interface Card}
  \acro{NTP}{Network Time Protocol}
  \acro{MAC}{Media Access Control}
  \acro{OS}{Operating System}
  \acro{OUI}{Organizationally Unique Identifier}
  \acro{SOHO}{Small Office-Home Office}
  \acro{U/L}{Universal/Local}
  \acro{SLAAC}{Stateless Address Autoconfiguration}
  \acro{VPS}{Virtual Private Server}
\end{acronym}

\end{document}

%% file: preamble.tex
\usepackage[subtle]{savetrees}
\usepackage{fancyvrb}
\usepackage{algpseudocode}
\usepackage{cleveref}
\usepackage{tabularx}

\usepackage[nolist,nohyperlinks]{acronym}
\usepackage{color}
\usepackage{graphicx}
\graphicspath{{./figs/}}
\usepackage[labelformat=simple,labelsep=space]{subcaption}
\usepackage{xspace}
\usepackage{multirow}
\usepackage{makecell}
\usepackage[ruled,vlined]{algorithm2e}

\usepackage{ulem}
\newcommand{\etal}{et~al.\xspace}
\newcommand{\vfour}{IPv4\xspace}
\newcommand{\vsix}{IPv6\xspace}
\newcommand{\eui}{EUI-64\xspace}

\newcommand{\zmap}{ZMap\xspace}
\newcommand{\zgrab}{ZGrab2\xspace}

\newcommand{\zmapsix}{ZMap6\xspace}
\newcommand{\icmpvsix}{ICMPv6\xspace}

\newcommand{\wifi}{Wi-Fi\xspace}
\newcommand{\totalntpnets}{13,601,343\xspace}
\newcommand{\totalcablenets}{8,799,019\xspace}
\newcommand{\totalscannets}{2,252,548,864\xspace}

\newcommand{\totalicmpresp}{93,832,220\xspace}

\newcommand{\eg}{e.g., \@}

\newcommand{\parhead}[1]{\medskip \noindent \textbf{#1}\hskip .1in}
\newcommand{\Parhead}[1]{\noindent \textbf{#1}\hskip .1in}

\ifdefined\isFinalized
\newcommand{\NewCommentType}[3]{}
\else
\newcommand{\NewCommentType}[3]{\expandafter\newcommand\csname #1\endcsname[1]{{\color{#2}{#3: ##1}} }}
\fi

\usepackage[override]{cmtt}

%% file: dml-abstract.tex
IPv4 NAT has limited the spread of IoT botnets considerably by default-denying
bots' incoming connection requests to in-home devices unless the owner
has explicitly allowed them.
As the Internet transitions to majority IPv6, however, residential connections no longer
require the use of NAT.
This paper therefore asks: has the transition from IPv4 to IPv6 ultimately made
residential networks more vulnerable to attack, thereby empowering the next
generation of IPv6-based IoT botnets?

To answer this question, we introduce a large-scale IPv6 scanning methodology
that, unlike those that rely on computationally intensive algorithms, can be run on low-resource devices common
in IoT botnets.
We use this methodology to perform the largest-scale measurement of IPv6
residential networks to date, and compare which devices are publicly accessible
to comparable IPv4 networks.
We received responses from 93,832,220 distinct IPv6 addresses, 14,901,892 of
which are inside of residential networks (i.e., not the external-facing
gateway IP). These residential network internal addresses span 4,412 ASes across
144 countries.
These responses come from protocols commonly exploited by IoT botnets
(including telnet and FTP), as well as protocols typically associated with
end-user devices (including iPhone-Sync and IPP).
Compared with IPv4, we reach 11$\times$ more HP printers over IPv6 than Shodan
finds across the entire IPv4 Internet, and thousands of iPhones and smart
lights whose services IPv4 NAT made unreachable by default.
Collectively, our results show that NAT has indeed acted as the de facto
firewall of the Internet, and the IPv4-to-IPv6 transition of residential networks
is opening up new devices to attack.
Finally, we discuss potential mitigations to prevent in-home network
reachability as \vsix adoption continues to grow.

%% file: intro.tex
\section{Introduction}
\label{sec:intro}

Residential networks are of paramount concern for Internet security.
Users all over the world increasingly deploy vulnerable
smart-home devices~\cite{hajime, iot-penetration-2024}.
IoT botnets have taken advantage of vulnerable in-home devices to launch
unprecedentedly large
attacks~\cite{antonakakis2017understanding,krebs-attack-2025}, thereby
threatening not only the device owners' own security and privacy, but
everyone on the Internet.

To date, IoT botnets' growth has been limited by a small, unlikely
defender: IPv4 Network Address Translation (NAT).
Because IPv4 addresses are in limited supply, almost all IPv4
residential networks translate addresses at their gateway, between the
home's single public IPv4 address and its many internal private
addresses.
One of NAT's accidental benefits is that it acts as a sort of firewall:
it default-denies incoming connections unless a user has explicitly set up a
port-forwarding rule to permit them.
IoT botnets have therefore been limited in their efforts to scan
devices on IPv4 residential networks, targeting almost exclusively
gateway routers (which have a public IP address) and webcams (which
commonly have NAT-bypassing port-forwarding rules).
A litany of other vulnerable smart-home devices has been largely
unreachable behind \vfour NAT.

IPv6, conversely, has a virtually limitless number of public addresses,
eliminating the need for NAT.
As a result, when residential networks enable IPv6, they lose an accidental
defense against IoT botnets.

There are two defensive mechanisms that \emph{should} fill the inadvertent gap
NAT leaves.
First, home gateway devices should deploy \emph{actual}
firewalls.
Second, IPv6's astronomically large address space (128 bits) is
expected to render scanning ineffective.
Some have theorized that these defenses are not present on the \vsix Internet
(see \S\ref{sec:background}), but to date there have been no empirical
studies, and moreover there are no established methods for effectively
scanning residential IPv6 networks to perform such a study.

It therefore remains an open question whether the transition from IPv4
to IPv6 has ultimately made devices inside of residential networks more
vulnerable to attack by IPv6-based IoT botnets.

We show in this paper that it is in fact possible to perform
large-scale scans of IPv6 residential networks in a manner that
low-resource devices common in IoT botnets can execute.
We perform the largest measurement of IPv6 residential networks to
date, and compare which devices are publicly accessible to comparable
IPv4 networks.
We present two key findings:

\parhead{Effective, low-computation IPv6 scanning}
We introduce a scanning methodology that is computationally low-cost
yet highly effective.
Given a known-active IPv6 /48 prefix (such prefixes are publicly available),
our technique simply scans the initial low-order addresses
(\texttt{::1}, \texttt{::2}, etc.) within each /56 prefix.
Though this is a narrower portion of the IPv6 space than what
compute-intensive target generation algorithms (\S\ref{sec:background}) could
potentially access, we show that it still allows us to
reach over ten million in-home devices.
Most importantly, it does so in a way that would be easily deployable
on any low-resource IoT device.

Our technique thus shows that scanning large portions of the
residential IPv6 Internet is within reach of IoT botnets.

\parhead{An analysis of vulnerable services publicly accessible over IPv6}
We applied our scanning methodology to perform the most comprehensive
measurement study of residential IPv6 networks to date.
We scanned for the presence of 30 port/protocol combinations, including
many of those most commonly used by IoT
botnets~\cite{antonakakis2017understanding,hajime}, as well as the
\vsix iPhone-Sync protocol (to our knowledge, this is the first measurement
of the public accessibility of iPhones).
We find millions of publicly accessible, internal residential IPv6
hosts running potentially vulnerable services.
We also compare our IPv6 findings to IPv4 through Shodan's scans of the
\emph{entire} IPv4 address space: for HP printers we reach an order of
magnitude more devices over \vsix than exist in all of \vfour, while for
services such as iPhone-Sync and smart lighting---which IPv4 NAT renders
unreachable altogether---\vsix reachability is net-new exposure.

Collectively, our results show that \vsix deployment in
residential networks is opening up new devices to attack, empowering a
new generation of IoT botnets, and ultimately putting the entire
Internet at greater risk.

\parhead{Contributions}
Our primary contributions are as follows:

\begin{itemize}

\item We introduce a scanning methodology that is significantly simpler
	than prior work, thereby demonstrating that even a low-resource IoT
	botnet could propagate across the IPv6
		Internet~(\S\ref{sec:methodology}).

\item We perform the most comprehensive scan of devices within \vsix
	residential networks, measuring the public accessibility of 30
	distinct port/protocol combinations commonly used by IoT botnets,
		and perform the first measurement of the reachability of Apple iOS
		devices.

\item We compare reachability across IPv4 and \vsix: we reach 11$\times$
	more HP printers than Shodan finds in all of IPv4, and iPhones, lighting
	systems, and IP cameras---unreachable through IPv4 NAT---by the hundreds
	to thousands~(\S\ref{sec:results} and \S\ref{sec:casestudies}).

 \item We propose potential remediation based on our measurements and
	 findings~(\S\ref{sec:mitigations}).

\end{itemize}

To facilitate further study and to encourage greater adoption of
defenses, we make our code and data available to the extent
that is safe and ethical~(Appendices~\ref{sec:open-science} and~\ref{sec:ethics}).

%% file: background.tex
\section{Background and Related Work}
\label{sec:background}

The core question of this paper is: Could IoT botnets today connect to millions
of hosts within residential IPv6 networks?
In this section, we review the IoT botnet threat model, give background on
IPv6, and present prior work in IPv6 scanning.

\subsection{Threat Model: IoT Botnets}
\label{subsec:iot}

Our threat model is meant to reflect the capabilities of today's IoT botnets
that target devices primarily located within residential networks.
The central goal of these adversarial bots is to connect to and exploit as many
vulnerable devices as possible.
To achieve this, the bots scan IP addresses.

\parhead{How IoT botnets target IPv4 hosts}
To date, all IoT botnets of which we are aware have operated solely over IPv4.
Specifically, they generate random IP addresses and attempt to connect to them
over a set of ports corresponding to services against which the bot has
exploits.
This has proven effective over
IPv4~\cite{antonakakis2017understanding}---because it comprises only 4B
addresses---but enumerating the entire IPv6 address space would, at a 10,000
packets per second probing rate, require more than $10^{27}$ years.
For IoT botnets to effectively scan IPv6, a new scanning methodology is
necessary, which we provide.

\parhead{Low-resource devices}
IoT botnets comprise primarily devices with low computation and memory resources,
especially so-called ``smart home'' devices that have infamously poor security,
such as default passwords and well-known
CVEs~\cite{antonakakis2017understanding,hajime}.
Resource constraints have no implications for how IoT botnets scan IPv4 today,
but as we discuss later in this section, they preclude them from running the
state of the art in IPv6 scanning (TGAs).

\parhead{Lists of active prefixes}
We assume that the adversary can obtain a public dataset of IPv6 /48 prefixes
that each have at least one active (though not necessarily reachable) host.
These lists are readily available~\cite{observatory}, relatively small when
compressed, and could be easily disseminated on today's IoT botnet
command-and-control infrastructure~\cite{hajime}.
Starting with these prefixes, the adversary's goal is to discover full /128
IPv6 addresses belonging to active, reachable hosts within residential
networks, as that is where vulnerable smart-home devices are likely to be.

We limit our adversary to knowing only active /48s because that is the most
that we authors can ethically use (see Appendix~\ref{sec:ethics}).
In practice, an adversary could participate in running a popular service and
scan back the full /128s of the clients that come back to
them~\cite{time-to-scan}.
Even then, the adversary could \emph{also} apply our techniques that derive
/128s from known-active /48s.
Thus, our attacker is strictly weaker than an attacker in practice could be,
and yet we show that we are nonetheless able to connect to millions of devices
inside of residential IPv6 networks.

\subsection{Residential IPv6 Background}

IPv6 is the successor to IPv4; it uses an impossible-to-enumerate 128-bit
address space instead of \vfour's enumerable 32-bit space.
This increase also affects how addresses
are assigned, which we review here.

\begin{figure}[t]
    \centering
    \includegraphics[width=\linewidth]{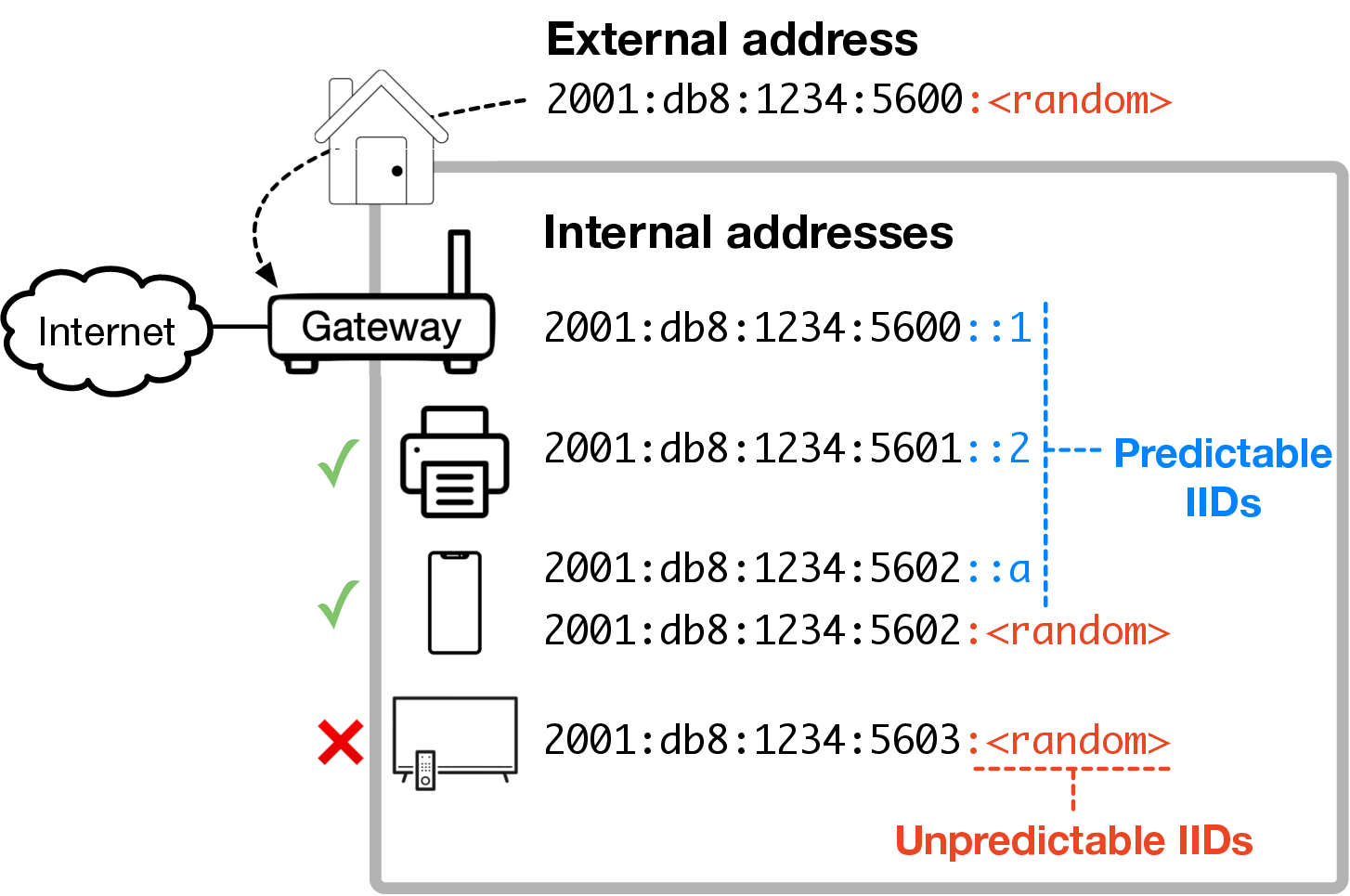}
    \Description{Diagram of a prototypical residential IPv6 network: a gateway router with an external address, and internal devices including a phone that holds both a random SLAAC address and a predictable low-order DHCPv6 address.}
	\caption{A prototypical residential network. Our methodology
	discovers devices with predictable IIDs, even if they also have
	unpredictable IIDs, like the phone in this example.}
    \label{fig:home}
\end{figure}

\parhead{Residential networks}
Residential networks, broadly speaking, consist of a gateway router
(CPE, or Customer Premises Equipment) that sits at the periphery of the
network, and a set of devices located \emph{inside} the network.
Figure~\ref{fig:home} provides a prototypical example of a residential network.

\parhead{IPv6 address assignments in residential networks}
In IPv4, residential networks typically get a single public IPv4 address, and
their gateways---running DHCP and NAT---assign a single private IPv4 address to
each device within their network.
Conversely, in IPv6, residential networks get an entire prefix from
their ISP, ranging in practice from a /48 to a /64~\cite{lacnic-prefix-delegation}.
Residential IPv6 gateways can then assign devices within their network a single
IPv6 address (via DHCPv6) or allow devices to self-generate \vsix addresses via
\ac{SLAAC} by providing them with a /64 (the hosts then choose their own lower
64 bits, known as an interface identifier (IID)).
A subtle but relevant detail is that devices may receive a /128 address assignment
via DHCPv6 \emph{but also} auto-generate addresses via \ac{SLAAC}.
We are the first to report on the prevalence of this dual-use
of DHCPv6, and we are the first to identify it as a security risk.

Unlike \vfour, in-home \vsix addresses are almost always
\emph{publicly routable}, meaning that unsolicited traffic can be sent to
residential LAN devices.

\parhead{Unpredictable IIDs}
IIDs are the lower 64 bits of an IPv6 address.
For routers and some public servers, network administrators commonly
choose predictable and low-entropy IIDs (e.g., \texttt{::1}), as they
are easier to remember and manually type~\cite{rfc7707}.
Predictable addresses can make it possible to scan~\cite{rfc7707},
track~\cite{rye2023hitlists,gasserapple}, and even geolocate users'
devices~\cite{ipvseeyou}.
Therefore, a decades-long best practice in IPv6 is for clients to
generate random IIDs and change them regularly, a process known as \ac{SLAAC}
Privacy Extensions or SLAAC-PE~\cite{rfc4941}.
Similarly, RFC~7707~\cite{rfc7707} recommends that, when DHCPv6 is used
for assigning full /128s, it should choose random addresses from a
large pool.
We are, to the best of our knowledge, the first to show empirically that
predictable addresses are \emph{still} being assigned to devices, or assigned
\emph{in addition} to randomly generated addresses.

\subsection{Scanning Residential IPv6 Networks}
\label{subsec:hard}

Unlike IPv4, IPv6's astronomically large address space makes it impossible to
scan every address.
Instead, IPv6 scanning techniques try to probe only addresses that are likely
to respond.
The central technical challenge behind IPv6 scanning is identifying IPv6
addresses that are likely to be in use.

\parhead{IPv6 hitlists}
Several efforts have sought to catalog \emph{IPv6 hitlists}: lists of
IPv6 addresses or prefixes that have been observed to be in use.
There are two noteworthy approaches:
The IPv6 Hitlist~\cite{expanse,hitlistwebsite} is collected via \emph{active}
scans of the Internet that elicit responses from live IPv6 devices.
Alternatively, Rye and Levin~\cite{rye2023hitlists} described a \emph{passive}
approach: running servers in the NTP Pool---a volunteer-run group of NTP
servers---and allowing IPv6 devices to come to them.

Rye and Levin compared these two approaches, finding that the IPv6 Hitlist
comprises mostly routers and public (non-residential) servers with low-entropy
IIDs, while the NTP-based hitlist can yield orders of magnitude more clients
with high-entropy IIDs.
However, even the NTP Pool-based hitlist is not comprehensive, as not all
devices make use of the NTP Pool (Apple and Android devices, for instance, use
their own NTP, non-NTP Pool time servers).
As the NTP-based approach yields more clients, we adopt it to seed our
scanning.
We will demonstrate that our methods can discover devices that never come to
NTP Pool servers, including Apple iPhones.

\parhead{Target generation algorithms (TGAs)}
TGAs are algorithms that, when trained on IPv6 data (typically previously active
addresses), generate new target IPv6 addresses to scan.
While there have been many distinct TGAs introduced over the past
decade~\cite{Foremski-entropy,6tree,det,6gcvae,6veclm,6gan,6hit,6scan,sixsense,6loda},
they all share two features that are relevant to our work:
\emph{First}, none of them can guess clients' randomly generated IIDs any
better than brute force.
\emph{Second}, they are all far too expensive in terms of computation and
memory to be run on IoT devices.
For example, 6Sense~\cite{sixsense} requires 512GB RAM/48GB GPU-RAM.
Today's TGAs are therefore not practical for IoT botnets to use for generating
scanning targets.
We present an effective, heuristics-based scanning methodology that is
sufficiently low-cost for virtually any IoT device to run.

\parhead{Prior studies of residential IPv6 gateways}
To the best of our knowledge, all prior scanning measurements of IPv6
residential networks have only studied the gateway routers sitting at
the networks' periphery.

Olson et al.~\cite{olson2021natting} studied ten gateway routers in a lab
setting and analyzed their security settings and firewall implementations.
They found that, contrary to RFC~3904's recommendations~\cite{rfc3904}, most
gateway routers do not enable a firewall by default.
Both Olson~\etal and RFC~3904 \emph{hypothesized} that IPv4 NAT has acted as a
de facto firewall, and that the transition to IPv6 could open new
attacks.
To the best of our knowledge, ours is the first work to demonstrate this
empirically, at scale, in the wild.

Several studies have identified and elicited responses from residential
gateways~\cite{edgy,li2021fast,czyz2016don,caida-ark}.
These generally work by sending packets to IPv6 addresses that are within a
residence's assigned network but are \emph{not} in active use, thus resulting in an
ICMPv6 Destination Unreachable message from the gateway, thereby revealing
the gateway's address.
These techniques enable scanning of the periphery of residential
networks, but they are not effective at reaching devices inside the
residential network.

\parhead{Prior studies of devices inside residential IPv6 networks}
Devices \emph{inside} of residential IPv6 networks are critical targets for IoT
botnets.
Yet, we are aware of no prior work that has scanned these devices at
significant scale.
The dearth of such studies is likely due to the difficulty in guessing
randomly generated IIDs: there are no known techniques for effectively scanning
devices inside residential networks at scale.

Rye and Levin~\cite{rye2023hitlists} learned of many IPv6 addresses within
residential networks, but they never scanned them, for various ethical
reasons we describe and account for in Appendix~\ref{sec:ethics}.

Some broader scanning studies that did not specifically target residential
networks have \emph{incidentally} detected hosts that may have been inside
residential networks.
Williams et al.~\cite{sixsense} used the 6Sense TGA
to scan IPv6 addresses broadly (not just residential networks), and
reported finding some devices commonly found in homes (e.g., printers).
Klopsch~\etal~\cite{time-to-scan} ran NTP Pool servers and scanned back the
addresses that came to them. This act has been prohibited by the NTP Pool
maintainers~\cite{ntp-pool-shodan-concerns}
(Appendix~\ref{sec:ethics}), and it misses the many such devices (e.g., iPhones)
that never use the NTP Pool. Our methodology is able to partially discover
these types of non-NTP Pool clients.

To the best of our knowledge, we are the first to scan devices inside residential IPv6
networks at scale: we elicited responses from millions of distinct devices from
within residential networks.
Moreover, our techniques can be run on resource-constrained devices commonly
present in IoT botnets.
Despite these constraints, and even though we limit our scans only to
residential networks, our techniques discover \eg over 128$\times$ more
printers than 6Sense~\cite{sixsense}.
In short, we show for the first time that it is possible for IoT botnets to
spread to the highly vulnerable devices that were long hidden behind NAT.

%% file: methodology.tex
\section{Scanning Methodology}
\label{sec:methodology}

\begin{figure}[t]
    \centering
    \includegraphics[width=\linewidth]{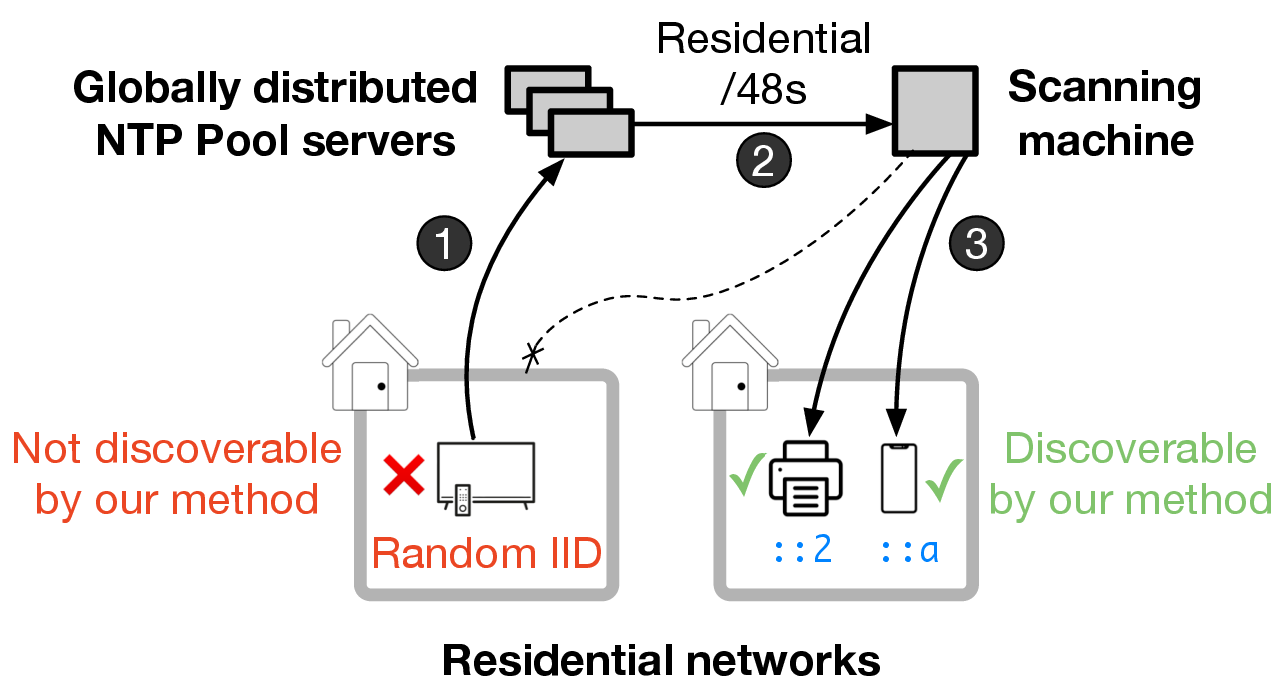}
    \Description{Three-step diagram: a client from an active /48 contacts one of our NTP Pool servers; the residential /48s are filtered out; the ::1 through ::a IIDs of every /56 in each residential /48 are probed, reaching devices that never contacted the NTP servers.}
	\caption{Overview of our scanning methodology. (1)~We learn of an
	active /48 when a client visits one of our NTP Pool servers. (2)~We
	extract only the residential /48s. (3)~We scan the
	\texttt{::1}--\texttt{::a} IIDs within each residential /56. This
	allows us to discover devices and networks even if they did not
	come to our NTP servers, like the network on the right in this
	example.}
    \label{fig:scanning-overview}
\end{figure}

In this section, we present our technique for scanning
devices inside of residential IPv6 networks at scale.
Figure~\ref{fig:scanning-overview} provides a high-level overview.
We then present how we determine whether a responses came from the external
interface of a gateway router or from and internal address.
Finally, we discuss limitations of our technique.

\subsection{Learning In-Use /48s via the NTP Pool}
\label{subsec:48s}

We learn active IPv6 addresses by running servers in the NTP
Pool~\cite{ntppool}, following the methodology of Rye and
Levin~\cite{rye2023hitlists}.
We ran 34 NTP servers in 25 countries (listed in Appendix~\ref{sec:ntp-pool-servers}) between July 2024
and January 2026.
The NTP Pool uses coarse IP geolocation data to direct NTP clients to a nearby
NTP server.
Thus, we obtain traffic from \emph{outside} the 25 countries our servers
are located in, as well.
Collecting active IPv6 addresses by running NTP Pool servers was shown by Rye
and Levin~\cite{rye2023hitlists} to result in far more client addresses than
other methods (\S\ref{sec:background}).

Our seed data comprised \totalntpnets distinct /48s in total.

\parhead{Why not just scan back the NTP Pool clients?}
At this point, it would be trivial to simply scan back the IPv6 addresses
that came to our NTP Pool servers.
However, prior to running our servers, we asked the NTP Pool operators
what forms of data they were comfortable allowing us to collect, share,
and act on.

The NTP Pool operators told us that it would \emph{not} be acceptable to scan
back the /128s due to their concerns that doing so would reduce user trust in
the project.
A decade ago, Shodan ran NTP Pool servers to scan back IPv6 addresses, which
drew widespread condemnation from the community~\cite{ntp-pool-shodan-concerns}.
In response, the NTP Pool administrators removed them. Other recent work has sought to
identify actors using the NTP Pool to learn IPv6 addresses with the aim of
restricting their future participation in the NTP Pool~\cite{rye2026cleaning}.

However, the NTP Pool operators also told us that sharing the /48s publicly was
acceptable (as with Rye and Levin~\cite{rye2023hitlists}), and that any
methodology that starts with \emph{only} the /48s and derives a /128 to scan
was also acceptable.
We therefore discard the lower 80 bits altogether from every client IPv6
address that comes to our NTP Pool servers, saving only the /48s.
The methodology we describe in the rest of this section starts with these /48
prefixes and derives /128s to scan.

\parhead{Alternatives open to botnets}
Botnet operators likely would not adhere to the same ethical standards we do;
they could run NTP Pool servers (or another popular service) and simply
scan back the /128s that come to them.
They could \emph{also} perform our techniques, which start with the /48s and
derive new /128s.
In that sense, our results are a strict lower bound on the set of devices a
botnet could potentially discover.

Alternatively, rather than run their own NTP Pool servers---which might raise
suspicion---a botnet could simply download public datasets of active
/48s~\cite{observatory}.
This would effectively be equivalent to the methodology we present in this
paper.

\subsection{Narrowing to Residential /48s}
\label{subsec:residential}
Our study focuses on \emph{residential} networks, but the NTP Pool
data also contains many non-residential /48s, including infrastructure
and mobile networks.

To focus on /48s allocated for residential networks, we keep only those that
the MaxMind GeoIP2 Connection Type database~\cite{maxmindconnection}
classifies as ``cable/DSL'' or ``dialup.''
This reduced the candidate set to 8,799,019 /48s (65\% of the original
13,601,343).
Applying this filter provides a degree of confidence that the
resulting /48 networks are assigned to residential or small business
networks by access ISPs.
We refer to these networks hereafter as ``residential,'' with the caveat
that MaxMind's cable/DSL category also covers small office/home office
(SOHO) networks served by the same access ISPs.

\subsection{Scanning Hosts within Residential /48s}
\label{subsec:scans}

Given residential /48s, we perform a series of active measurements to
identify and then scan responsive addresses within those prefixes.

\parhead{Address selection}
For a given /48, we iterate over each of its 256 /56s, and for each one
we send \icmpvsix Echo Requests to its first ten IIDs (\texttt{::1}--\texttt{::a}).
Unlike TGA-based IPv6 scanning techniques (\S\ref{sec:background}),
this simple heuristic requires no sophisticated computation and could
run on virtually any IoT device.

To evaluate how many low-value IIDs to use, we randomly sampled 1,000
/48s, scanned the first 1,000 IIDs of each /56, and recorded which ones
responded.
Figure~\ref{fig:mini-experiment} in the Appendix shows that the vast
majority (87.5\%) of responsive addresses fell within the first ten.

The intuition behind scanning low-value IIDs is that it ought to
capture IIDs that users set manually, as well as those given out by
DHCPv6 servers contrary to the best practices of RFC~7707~\cite{rfc7707}.
As we will see, the prevalence of DHCPv6 makes this a surprisingly
effective way of finding responsive devices---even devices that
\emph{also} assign themselves random IIDs.

\parhead{Detecting responsiveness and aliases}
For each address we consider for scanning, we first test its
responsiveness by sending an ICMPv6 Echo Request (ping).
We do this using \zmapsix~\cite{zmap6}, the \vsix port of
\zmap~\cite{durumeric2013zmap}.
If we do not receive any responses to this Echo Request, then it is
less likely that there is an active device at that address (although it is
possible that the device does not respond to \icmpvsix but does to other
protocols).
To limit the impact of our scans on networks, we do not subject those
addresses to subsequent port scans.
Botnets may not be so conservative, and would be able to probe hosts
that are unresponsive to Echo Requests.

We also send an Echo Request to a random address in each of the /56s we
probe.
If we receive a response from this randomly chosen address, we conclude
that the /56 network is \emph{aliased}: i.e., that it responds to
probes to \emph{any} address.
To avoid over-counting the number of reachable networks, we simply
remove aliased networks from our further scanning and analyses
altogether.

\parhead{Protocol scans}
After identifying responsive, non-aliased addresses, we use \zgrab to
scan them over 30 port/protocol combinations, including CWMP (router and VoIP
management), HTTP, telnet, SSH, IPP (Internet Printing Protocol), and MSSQL.
Descriptions of the port and protocol combinations we probed are given
in Appendix~\ref{sec:extra-tables}.
We selected these particular protocols due to their general popularity
as well as their being common targets of IoT
botnets~\cite{hajime,iotpot}.

To further evaluate the extent to which IPv6-based scans can better
reach traditionally NATted devices, we also perform scans of the
``iPhone-Sync'' and Android Debug Bridge (ADB) protocols, which are used by
iPhones and iPads to wirelessly sync with a computer over port
62078~\cite{usbmuxdapple} and to connect to Android devices, respectively.
For the iPhone-Sync protocol, we wrote a custom \zgrab module that simply
requests the iPhone's iOS version and then gracefully disconnects; it does not
leak any PII.
This module is publicly available~(Appendix~\ref{sec:open-science}).
To the best of our knowledge, we are the first to measure the
reachability of this protocol and the iOS versions deployed in the wild.

\subsection{Differentiating Responses from Internal vs.~External Addresses}
\label{subsec:classify}

Residential networks, like the example in Figure~\ref{fig:home},
consist of one or more \emph{external} addresses belonging to the
gateway router and a set of \emph{internal} addresses belonging to
both the gateway and, critically, devices inside the network.
Prior work has established how to elicit responses from external
addresses of residential IPv6
networks~\cite{edgy,li2021fast,czyz2016don,caida-ark}.
The central goal of our work is to scan internal addresses, but it is
possible that some of the responses to our probes were from external
addresses.
Thus, a critical part of our post-processing is to differentiate the
two.

\parhead{Responses from external addresses}
Every response to an \icmpvsix \zmap probe can be matched to the probe that
elicited it.
If any response comes from an address different from the destination
to which we sent the probe, then we classify the responding address as an
external address.
For example, if the IPv6 address \texttt{2001:db8::928a:bac2:da01:19fc} responded to the ICMPv6
Echo Request that we sent to an address within a target /56, then we classify
that IP address as the external address for that /56.

Responses from these external addresses are typically
\icmpvsix Destination Unreachable type error messages (e.g.,
Administratively Prohibited or Address Unreachable).
In residential \vsix networks, messages of these types generally
originate from a gateway's external interface~\cite{edgy,rye2021follow}
when the gateway cannot route the traffic to the destination, either
because the destination address is not in use or its security policy
does not permit it.

\parhead{Responses from internal addresses}
If we receive an \icmpvsix Echo Reply from the address that we
probed, we classify it as an \emph{internal address}.
For example, if \texttt{2001:db8::1} responds to our ICMPv6
ping with an \icmpvsix Echo Reply, we classify it as an internal address for
that /56.
In the event that the same address is in both sets, we remove those networks
from further consideration, as we cannot differentiate between the internal
\vsix address space assigned to that home network, and the address assigned to
the external interface of the gateway device.
This affected 5,119,944 addresses (5.5\% of all responses), which we exclude
from the internal and external counts in \S\ref{sec:results}.

\subsection{Limitations}
\label{subsec:limitations}

\Parhead{Methodological limitations}
Our techniques have several inherent limitations that prevent them from
comprehensively scanning all residential IPv6 hosts:

First and most obviously, we only scan the first ten IIDs
(\texttt{::1}--\texttt{::a}), which of course not all active devices
make use of.
We trade off comprehensiveness for feasibility of computation on
resource-constrained devices, and show in \S\ref{sec:results},
surprisingly, that it suffices to reach millions of devices inside of
residential IPv6 networks.

Second, as discussed in~\S\ref{sec:background}, not all devices use the
NTP Pool, so it is possible that we are missing entire active /48s from
our initial set of prefixes to probe.
Although Windows, iOS/macOS, and some versions of Android do not use
the NTP Pool, many versions of Linux do, and many IoT devices use
Linux.
Note also that we do not need a device to come to our NTP Pool servers
in order for us to potentially probe it; if \emph{any} device
within its /48 comes to our NTP Pool servers, then we can potentially
scan it.
Thus, despite this limitation, we are still able to discover millions
of devices, including those that scanning back would
not~\cite{time-to-scan}.

Third, \vsix (like \vfour) can have highly dynamic addresses.
Some ISPs reassign residential prefixes as frequently as every 24
hours~\cite{dynamic,padmanabhan2020dynamips,rye2021follow}.
This complicates measurement campaigns longer than the prefix rotation
period, as probes sent to a network at one point in time may end up in
a different customer's network at a later point in time.
To minimize these effects, we tightly coupled address discovery using \zmapsix
with the application-layer \zgrab scans we conducted.

Fourth, in some networks, the gateway may be assigned an address
between \texttt{::1} and \texttt{::a} on its external interface.
For gateways that are configured this way, we will erroneously consider
the responsive \vsix address internal, rather than external.
We eliminate instances in which this happens, as it confounds differentiation
between the home network internal address space and the external address
assigned to the gateway's WAN interface.

\parhead{Self-imposed limitations}
In addition to the above limitations inherent to our methodology, we
also imposed three constraints on ourselves to minimize the potential
load on residential networks:

We scanned at the granularity of /56s, which many but not all ISPs
delegate to residences; some use finer granularities (especially /60 or
/64).
In such cases, we fail to probe every residential network.
One could apply our methodology at finer granularities, but with
greater likelihood of re-scanning homes that have larger prefixes.

We scanned from four cloud vantage points, all of which are located in the
United States.
A lack of vantage point diversity in IPv4 has been shown to miss a small but
significant fraction of HTTP(S) hosts (less than 10\%) and a moderate fraction
of SSH hosts (less than 20\%)~\cite{origin-of-scanning}.
We are unaware of a similar study over IPv6, but we also expect that we
are missing hosts we might have seen with additional vantage points,
because the root causes identified over IPv4 (loss, blocking, and
outages) likely translate over to IPv6 to some extent.
Note, however, that the NTP Pool servers used to seed our study were
geographically distributed across 25 countries (Appendix~\ref{sec:ntp-pool-servers}).

Finally, our scans use \icmpvsix responsiveness as a prerequisite for
subsequent scanning, but devices can be responsive to network services
without being reachable via (or responsive to) pings.
One could port-scan without first performing a responsiveness check,
but that would increase the likelihood of scanning unused addresses.

\parhead{Summary of data collection decisions}

We believe that the heuristics and methodological decisions we make
are reasonable, and that our results therefore constitute a lower bound on what
a botnet might discover in probing IPv6.
First, we seed our scanning from active NTP Pool /48s (\totalntpnets), so
networks with no Pool client are never probed and the client addresses
themselves are not probed. Second, we keep only the /48s MaxMind labels
cable/DSL or dialup (\totalcablenets, 65\%), which discards misclassified
residential prefixes. Next, we enumerate every /56 of each /48, so homes
delegated a /60 or /64 are probed at the wrong granularity or not at all, which
skips potentially reachable devices in these networks. We probe ten low-order
IIDs plus one random alias check per /56, so devices with predictable DHCPv6
addresses above \texttt{::a} are unprobed.  We port-scan only the addresses that
respond to ICMPv6 Echo Requests, which misses devices that may be unresponsive
to ICMPv6 but do respond to port scans.  Finally, we classify a reply from the
probed address as internal and an error from any other address as external,
dropping the 5,119,944 addresses that appear in both sets.

\parhead{Implications for results}
All of the above heuristics, methodological decisions, and limitations cause our experiments to
\emph{underestimate} the number of reachable hosts within residential
IPv6 networks.
We do not anticipate that our limitations bias our results towards any
particular networks or regions of the world, other than perhaps the
geographical bias of our vantage points in the United States.
Thus, we believe our results are representative (but a strict subset)
of what an actual IoT botnet could achieve.

It is not possible to evaluate precisely how much of an underestimate
our results are.
Rather than evaluate comprehensiveness, we view our results through a
more pragmatic lens: whether or not it is possible for an IoT botnet to
reach a significant number (millions) of hosts within residential
networks.

%% file: results.tex
\section{Results}
\label{sec:results}

In this section, we describe the results of our \vsix residential
network measurement campaign.

\subsection{Seed Data Coverage}

Because we limit our scans to the /48s we learn from running our NTP
Pool servers, the coverage, accuracy, and potential biases of our
results ultimately derive from that data.
We therefore begin this section by analyzing the representativeness of
our seed dataset.
To the best of our knowledge, this is the first analysis of strictly residential
networks learned from NTP Pool servers~\cite{rye2023hitlists}.

Of the \totalntpnets /48s observed from our 34 NTP servers in 25 countries,
we categorized 8,799,019 (65\%) as residential/SOHO networks
(\S\ref{subsec:residential}).
These prefixes originate from a wide range of \acp{AS} and countries:

\parhead{Coverage of countries}
Herwig et al.~\cite{hajime} observed that IoT botnets
infect some countries far more than others.
To reason about how the IPv4-to-IPv6 transition might impact the spread
of botnets, it is important to have broad coverage across many countries.

To evaluate our dataset's country coverage, we aggregated the seed
/48s by the country their AS is registered in, according to Team
Cymru's ASN lookup service~\cite{cymru}, which relies on
\texttt{whois}.
Table~\ref{tbl:country-coverage} shows the top 5 countries---which
collectively constitute about 75\% of the active residential /48s we
observed.
All of these have sizable \vsix deployments~\cite{googleipv6}.
Japan is the most common country, with slightly less than a quarter of
all residential /48s. The US and China also contribute more than 10\% of the
/48s.
However, we will see that, surprisingly, some of the countries with the
most publicly accessible residential hosts are not in this top-5.

\begin{table}[tb]
\centering
\caption{Number of active residential /48s per country, as observed
    from our NTP Pool servers.}
\label{tbl:country-coverage}
\begin{tabular}{l l r r}
  \toprule
  \textbf{Rank} & \textbf{Country} & \multicolumn{1}{c}{\textbf{/48s}} & \multicolumn{1}{c}{\textbf{\%}} \\
  \midrule
  1 & JP            & 2,071,493 & 23.54 \\
  2 & US            & 1,670,569 & 18.99 \\
  3 & CN            & 1,231,894 & 14.00 \\
  4 & DE            & 768,617 &  8.74 \\
  5 & BR            & 756,514 &  8.60 \\
  \midrule
  \multicolumn{2}{l}{Other (157 countries)} & 2,299,932 & 26.14 \\
  \midrule
  \multicolumn{2}{l}{Total} & 8,799,019 & 100.00 \\
  \bottomrule
\end{tabular}
\end{table}

\parhead{Coverage of ASes}
We aggregated the residential /48s by the AS announcing them using
contemporaneous Routeviews~\cite{rv} BGP RIB data from January 2026.
5,874 ASes are represented by at least one /48 in our data.
The most common AS is KDDI (AS2516), a Japanese ISP, with about 20\% of all
residential /48s.
An American, Chinese, German, and Norwegian ISP round out the top five.

Collectively, these coverage results indicate that our NTP Pool-derived
data contains a geographically diverse set of candidate
networks to probe.
Moreover, the most highly represented countries align with some of the most
highly infected countries of the Hajime IoT botnet~\cite{hajime}.
We do not claim that this is necessarily representative of the
\emph{entire} IPv6 residential Internet.
However, given that our scanning methodology is the first that would be
feasible for common IoT botnet devices to perform, we believe that our
results are representative of the coverage an IoT botnet could obtain.

\subsection{Which Residential IPv6 Networks Respond?}

Here, we evaluate how well our scanning methodology
(\S\ref{subsec:scans}) is able to discover responsive hosts inside of
residential networks.
Recall that our scanning methodology takes as input the 8.8M
known-active residential /48s observed from our NTP Pool servers and first sends an
ICMPv6 Echo Request (ping) to the first ten addresses
(\texttt{::1}--\texttt{::a}) and one random IID within each /56.
In total, this amounted to roughly 24B probes, sent at 10,000 packets
per second.
Recall further that we remove all aliased networks from consideration using
responsiveness to the random IID probe as an indicator of aliasing (since it is
extraordinarily unlikely that a randomly chosen address is assigned to
a live host).
We found only 58,888 /56 networks that were aliased, a small fraction
of the \totalscannets we probed (0.0026\%).

In total, \totalicmpresp unique addresses responded to our ICMPv6 probes,
after excluding aliased networks.
This includes both \icmpvsix Echo Replies as well as \icmpvsix error responses:
both indicate a responsive host that can be further scanned.
The responsive addresses span 2,414,088 /48s, 6,071 \acp{AS}, and 152
countries\footnote{We use the term ``countries'' throughout, though we
report ISO 3166-1 alpha-2 country codes, which include dependent
territories.}.

\parhead{Internal vs.~external responses}
We distinguish response addresses as being external (likely coming from
the CPE) or internal (likely coming from a device within the
residential network).
If the response contains an \icmpvsix error from an address we did not
probe, then we conclude the reply likely came from the home gateway's WAN-facing interface; we
classify those as external responses.
If it contains the address we probed for, then we classify it as an
internal response.

The majority of the responses we received, 73.8M (78.7\%), came from
external \vsix addresses.
These addresses originate from 5,314 \acp{AS} and 143 countries; using
Team Cymru's IP-to-ASN service~\cite{cymru}, the most common \acp{AS}
include Chinanet Backbone (AS4134) and China Mobile (AS9808), while
the most common countries these responses originate from are China, Brazil,
Germany, and Japan.

We received responses from 14.9M (15.9\%) distinct \emph{internal}
\vsix addresses.
These addresses originated from 4,412 distinct \acp{AS} and 144 countries.
The remaining 5,119,944 responsive addresses (5.5\%) appeared in both sets
and are excluded (\S\ref{subsec:classify}).
There is moderate overlap in the most common \acp{AS} and countries
with external addresses---the most common \acp{AS} include
Verizon (AS701), BT (AS2856) and Chinanet Backbone (AS4134). The most common
countries include the US, the UK, Brazil, and China.

\parhead{Which IIDs are most successful?}
We scanned the first ten IIDs (\texttt{::1}--\texttt{::a}) in each /56
that we probed.
Among internal responses, \texttt{::1} is 31$\times$ more likely to
respond than the second-most common IID, \texttt{::2}.
Figure~\ref{fig:internal-iid-bar} in the Appendix displays the distribution of
internal \vsix response IIDs.

We speculate that some CPE devices self-assign \texttt{::1} to their
LAN interface, leading it to be more common than other IIDs.
To test this hypothesis, we computed the hop limit (\vsix's analog to \vfour's
Time-to-Live (TTL)) distance from the
responsive hosts to our measurement machine from the \vsix response headers.
More precisely, we assumed a starting hop limit of 64 if the received hop limit
was less than 64, 128 if greater than or equal 64, and 255 if greater than or
equal to 128, given that OSes typically set this value to a power of 2 or the
maximum value of 255 when they create \vsix packets.
For example, if the response we received had a hop limit of 118, then
we assume it started at 128 and was thus a distance of 10 hops away.

For the 5,960,191 residential networks where we received both an internal
\emph{and} an external response (many networks had only one or the other), we
computed the difference in distances between them.
A difference of zero indicates that the external and internal
addresses correspond to interfaces on the same device; a positive
difference indicates a device behind the home gateway.

\begin{figure}[tb]
  \centering
    \includegraphics[width=\linewidth]{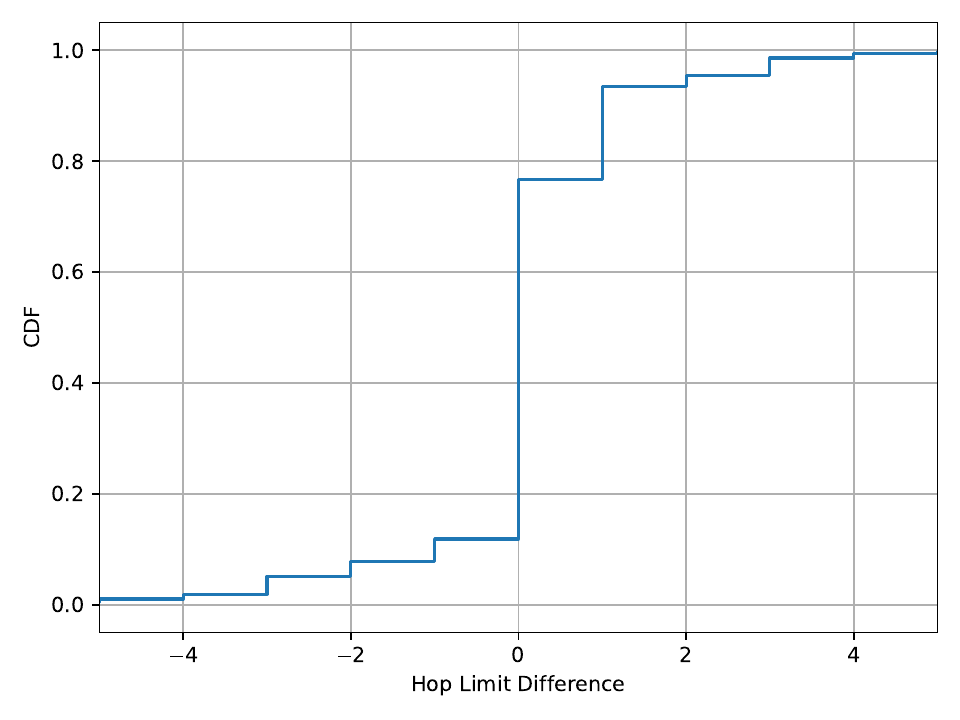}
    \Description{CDF of the hop-limit distance between paired internal and
    external addresses in the same /56, over 5,960,191 pairs; most are at distance
    zero, and about 12 percent are negative.}
        \caption{CDF of distance between internal and external hosts in the same
        /56 ($N=5,960,191$ pairs).}
      \label{fig:ttls}
\end{figure}

Figure~\ref{fig:ttls} shows the distribution of the differences in
distance between internal and external responses from the same network.
Indeed, we find that a difference of zero is the most common case,
constituting 3,863,772 (65\%) of the internal-external \vsix pairs that
responded to us.
Another 997,436 (17\%) of internal-external \vsix pairs are at a
distance of +1, indicating that the device assigned the internal \vsix
address is one hop behind the residential gateway.
Larger differences in hop limits occur in 390,950 cases (7\%), and may arise
from path changes, load balancing~\cite{global-censorship-routing}, or
an incorrect assumption about the starting hop limit.
The remaining 708,033 pairs (12\%) have a \emph{negative} difference. Potential
explanations for negative distances include topology changes or renumbering
during our experiment, middleboxes that rewrite hop limits between the probed
network and our vantage point~\cite{kappes2026ttl}, and nonstandard initial Hop
Limits.

\subsection{What Network Services Can Be Accessed?}
\label{subsec:protocol}

Having established which residential \vsix addresses are
\emph{responsive}, we next turn to evaluating which ones are running
services that are \emph{accessible}.
The last stage of our scanning measurement scans the 93.8M addresses
that responded to our ICMPv6 pings for 30 different
port/protocol pairs (\S\ref{subsec:scans}).
Here, we report on who responded, to what protocols, and how this
differs between external (CPE) and internal (LAN-side) responses.

\parhead{Overall accessibility}
Of the \totalicmpresp \icmpvsix-responsive addresses, 6,291,349 (7\%)
responded to at least one of the protocols we probed.
Table~\ref{tbl:responsiveness} shows how many distinct port/protocol
combinations
each of the responsive addresses responded to.
75.4\% of them respond to only a single service, while 24.6\% can be
reached via two or more.

\begin{table}[t]
\centering
\caption{IPs responding to exactly $k$ port/protocol combinations (cumulative \% is $\leq k$), of 6,291,349.}
\label{tbl:responsiveness}
\begin{tabular}{rrr}
\toprule
\textbf{Port/Protocols} & \textbf{IPs} & \textbf{Cumulative \%} \\
\midrule
1 & 4,741,151 & 75.4\% \\
2 & 886,665 & 89.5\% \\
3 & 433,011 & 96.3\% \\
4 & 164,398 & 98.9\% \\
5 & 41,759 & 99.6\% \\
6 & 21,289 & 100.0\% \\
7 & 2,824 & 100.0\% \\
8 & 173 & 100.0\% \\
9 & 67 & 100.0\% \\
10 & 10 & 100.0\% \\
11 & 0 & 100.0\% \\
12 & 2 & 100.0\% \\
\bottomrule
\end{tabular}
\end{table}

\parhead{External vs.~internal accessibility}
Internal \vsix addresses were slightly more responsive to protocol
scans, with 2,792,352 of 14,901,892 (18.7\%) responsive, compared to external
addresses' 3,453,701 of 73,810,384 (4.7\%) responsiveness.
This is a surprising result, given that 5$\times$ more external
addresses responded to our ICMPv6 pings than did internal addresses.
These numbers show that devices internal to a network are far more
likely to be running an openly accessible network service.

One possible explanation for this disparity is that network-internal
devices are---thanks to decades of NAT---tacitly assuming that any
machine accessing them must be coming from the same LAN.
Home gateway devices, on the other hand, have long understood the importance of
restricting access to their network services only to LAN-side devices.

\begin{figure}[t]
  \centering
  \includegraphics[width=.47\textwidth]{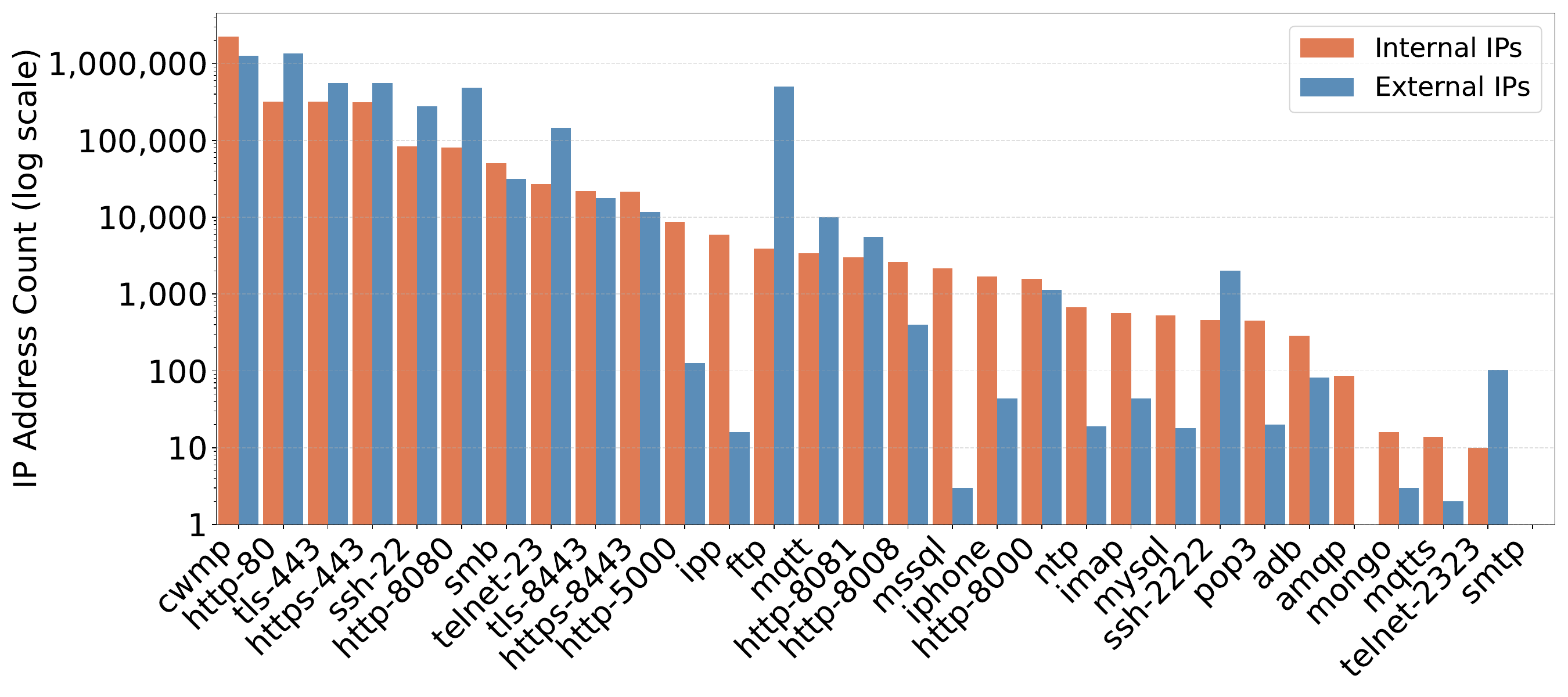}
  \Description{Bar chart with a log-scale y-axis of the number of responsive IPv6 addresses per protocol and port, with internal and external addresses shown as separate bars.}
  \caption{Overall number of responsive addresses by response type and
	protocol ($y$-axis logscale).}
	\label{fig:proto-bars-any}
\end{figure}

\parhead{Service accessibility}
We next compare the specific networked services that can be reached on
external and internal addresses.
Figure~\ref{fig:proto-bars-any} shows the number of responsive \vsix
addresses by protocol and response type.
Internal
addresses dominate protocols more likely to run on internal hosts, such as the
Internet Printing Protocol (IPP) and database protocols like MSSQL; external
addresses are more likely to be responsive to FTP and HTTP/80.

\begin{figure*}[t]
  \centering

    \begin{subfigure}[t]{0.48\linewidth}
      \includegraphics[width=\linewidth]{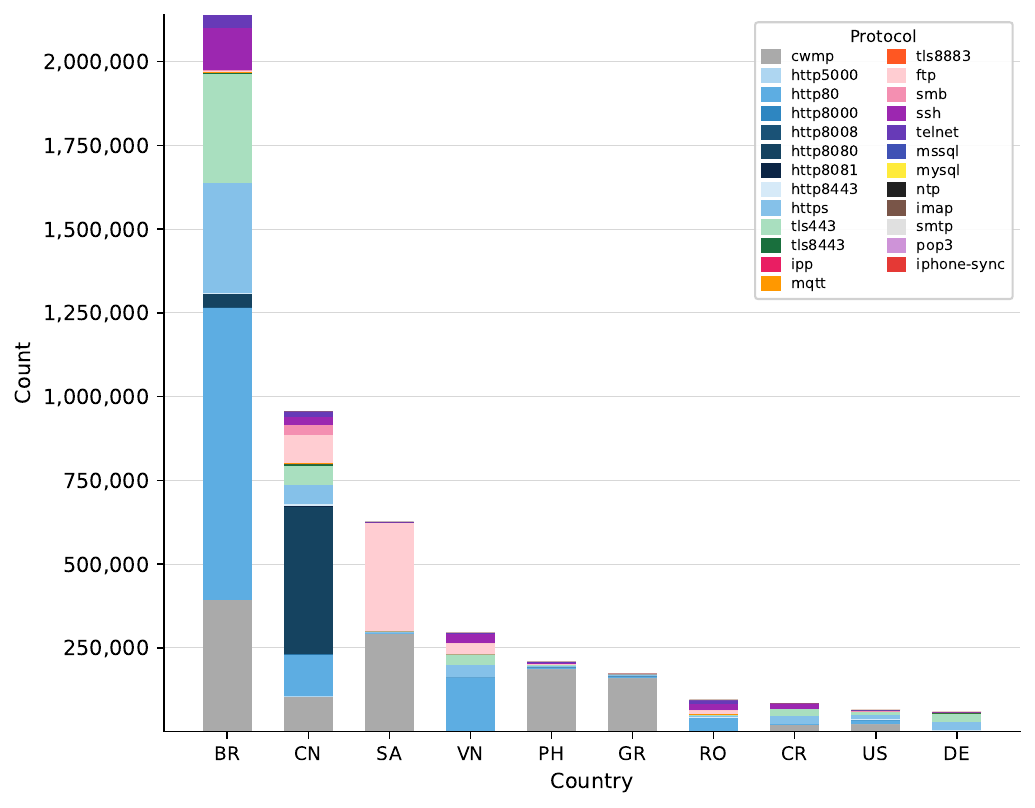}
      \Description{Stacked bar chart of the top ten countries by responsive external addresses, each bar broken down by network service.}
		\caption{Network services reachable on \emph{external} addresses.}
      \label{fig:external-stacked-bar}
    \end{subfigure}
    \hfill
    \begin{subfigure}[t]{0.48\linewidth}
      \includegraphics[width=\linewidth]{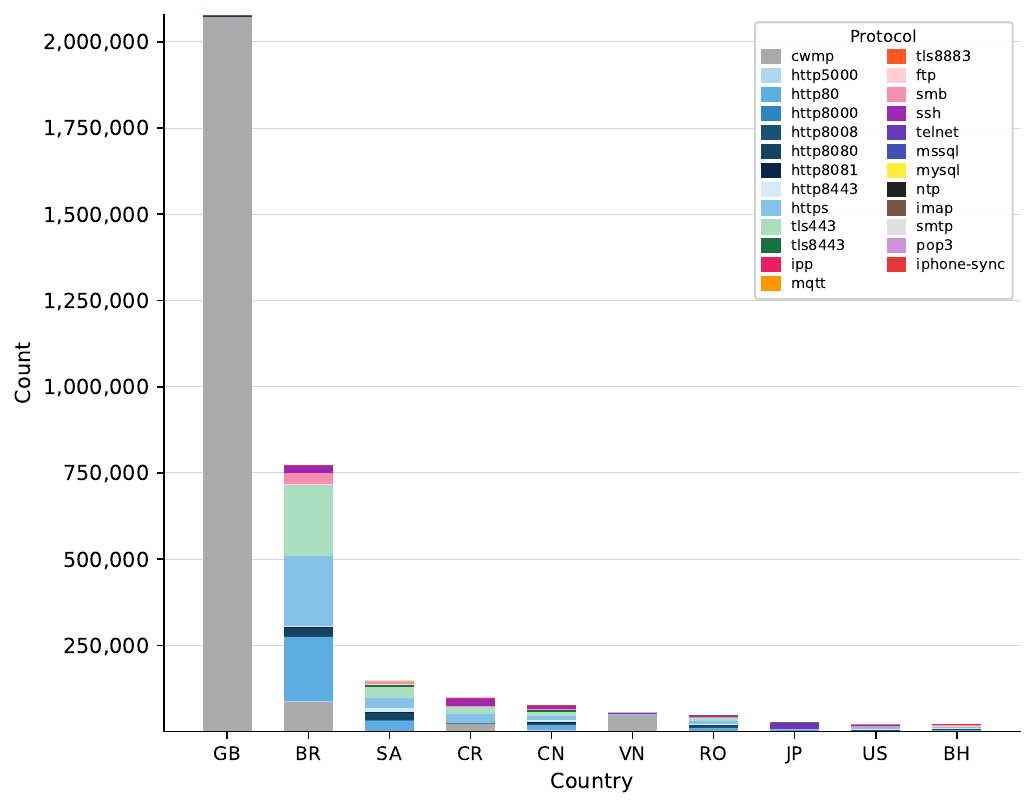}
      \Description{Stacked bar chart of the top ten countries by responsive internal addresses, each bar broken down by network service.}
		\caption{Network services reachable on \emph{internal}
		addresses.}
      \label{fig:internal-stacked-bar}
    \end{subfigure}

    \caption{The network services that are reachable on external IPv6
    addresses differ significantly from those reachable on internal addresses.}
	\label{fig:stacked-bar}
\end{figure*}

Figure~\ref{fig:stacked-bar} shows the top 10 countries with the most
responsive external (left) and internal (right) addresses, broken down
by the specific network services they responded to.
We make two key observations:

\emph{First}, internal and external accessibility can vary greatly by
country.
Note that the rank order of the top-10 (and even the set of countries
comprising the top-10) is not the same across both plots.

\emph{Second}, the set of services reachable via external addresses and
internal addresses differs significantly.
External addresses tend to host CWMP, web servers (HTTP and HTTPS), and
FTP---all services that IoT botnets are well-known to
attack~\cite{hajime}.
Internal addresses, on the other hand, host a much greater diversity of
services, including various database applications, IPP (printing), and
the iPhone-Sync protocol.
In Saudi Arabia (SA), for instance, external addresses tend to host CWMP and
FTP. However, internal addresses host a range of HTTP ports, potentially
indicating they host web pages to administer the gateway device on
internal-facing \vsix addresses and assume that it is restricted from the WAN.
This is a troubling finding, as it indicates that there are many more
potentially vulnerable services for future botnets to target.

\begin{table}[tb]
\caption{Protocols exposed on internal IPv6 addresses but not their paired external
  addresses, across all 10.8M same-network pairs.
  \emph{Pairs} counts (internal,~external) pairs with the service present only on
  the internal side; \emph{Sole diff.}\ counts pairs where that protocol is the
  \emph{only} asymmetric service; \emph{Unique internal IPs} counts distinct
  internal addresses with that exposure.}
\label{tab:internal-exposure}
\centering
\small
\begin{tabular}{lrrr}
\toprule
\textbf{Protocol (port)} & \textbf{Pairs} & \textbf{Sole diff.} & \textbf{Unique internal IPs} \\
\midrule
HTTP (80)               & 166{,}415 & 55{,}128 & 164{,}177 \\
TLS/HTTPS (443)         & 138{,}144 & 21{,}746 & 136{,}665 \\
HTTP (8080)             &  77{,}682 &  3{,}147 &  77{,}247 \\
SMB (445)               &  21{,}158 & 13{,}888 &  21{,}044 \\
SSH (22)                &  18{,}806 & 10{,}955 &  18{,}310 \\
TLS (8443)              &  13{,}640 &    789   &  13{,}521 \\
CWMP (7547)           &  10{,}514 &  8{,}328 &  10{,}336 \\
HTTP (5000)             &   6{,}172 &  5{,}548 &   6{,}095 \\
IPP (631)               &   5{,}567 &    809   &   5{,}527 \\
Telnet (23)             &   4{,}634 &  1{,}809 &   4{,}515 \\
HTTP (8008)             &   2{,}180 &    845   &   2{,}161 \\
MSSQL (1433)            &   2{,}053 &    146   &   2{,}042 \\
FTP (21)                &   1{,}942 &    127   &   1{,}864 \\
MQTT (1883)             &   1{,}669 &    852   &   1{,}652 \\
iPhone-Sync(62078)    &   1{,}512 &  1{,}493 &   1{,}489 \\
HTTP (8000)             &   1{,}317 &    345   &   1{,}312 \\
HTTP (8081)             &   1{,}159 &    512   &   1{,}134 \\
NTP (123)               &     646   &    331   &     644   \\
MySQL (3306)            &     436   &     62   &     433   \\
IMAP (143)              &     389   &     12   &     345   \\
POP3 (110)              &     321   &      0   &     282   \\
ADB (5555)              &     277   &    215   &     277   \\
SSH (2222)              &     118   &     43   &     105   \\
AMQP (5672)             &      78   &      6   &      78   \\
MongoDB (27017)         &      12   &      3   &      11   \\
MQTTS (8883)            &       9   &      1   &       9   \\
Telnet (2323)           &       3   &      2   &       3   \\
\midrule
\textbf{Total (any)}    & \textbf{257{,}712} & & \\
\bottomrule
\end{tabular}
\end{table}

\parhead{What devices allow unsolicited traffic?}
Networks with both internal and external responses provide an opportunity to
determine what types of CPE devices are allowing unsolicited traffic \emph{into}
residential LANs. Of 10,833,230 /56 networks with at least one \icmpvsix-responsive
internal address and one external address, 522,654 use \eui external \vsix addresses, embedding the
MAC address of the WAN interface into the lower 64 bits of the \vsix address.
These MAC addresses, when resolved against the IEEE's \ac{OUI}
database~\cite{oui}, are concentrated in a relatively small number of router
manufacturers. ZTE, eero, Taicang T\&W Electronics, China Mobile, and Shenzhen
Skyworth each have over
30k unique MAC addresses embedded in external \vsix addresses; ZTE leads all
manufacturers with over 126k unique MAC addresses.

The protocol scans of the external \vsix addresses also provide additional
context about which home router manufacturers permit unsolicited traffic into
residential LANs. For instance, we find over 111k unique external \vsix
addresses that present Nokia-issued TLS certificates and have reachable
internal \vsix addresses. More than 50k external \vsix addresses that permit
internal network reachability present TLS certificates issued by Intelbras, a
Brazilian network equipment manufacturer.

\begin{figure}[t]
  \centering
   \includegraphics[width=\columnwidth]{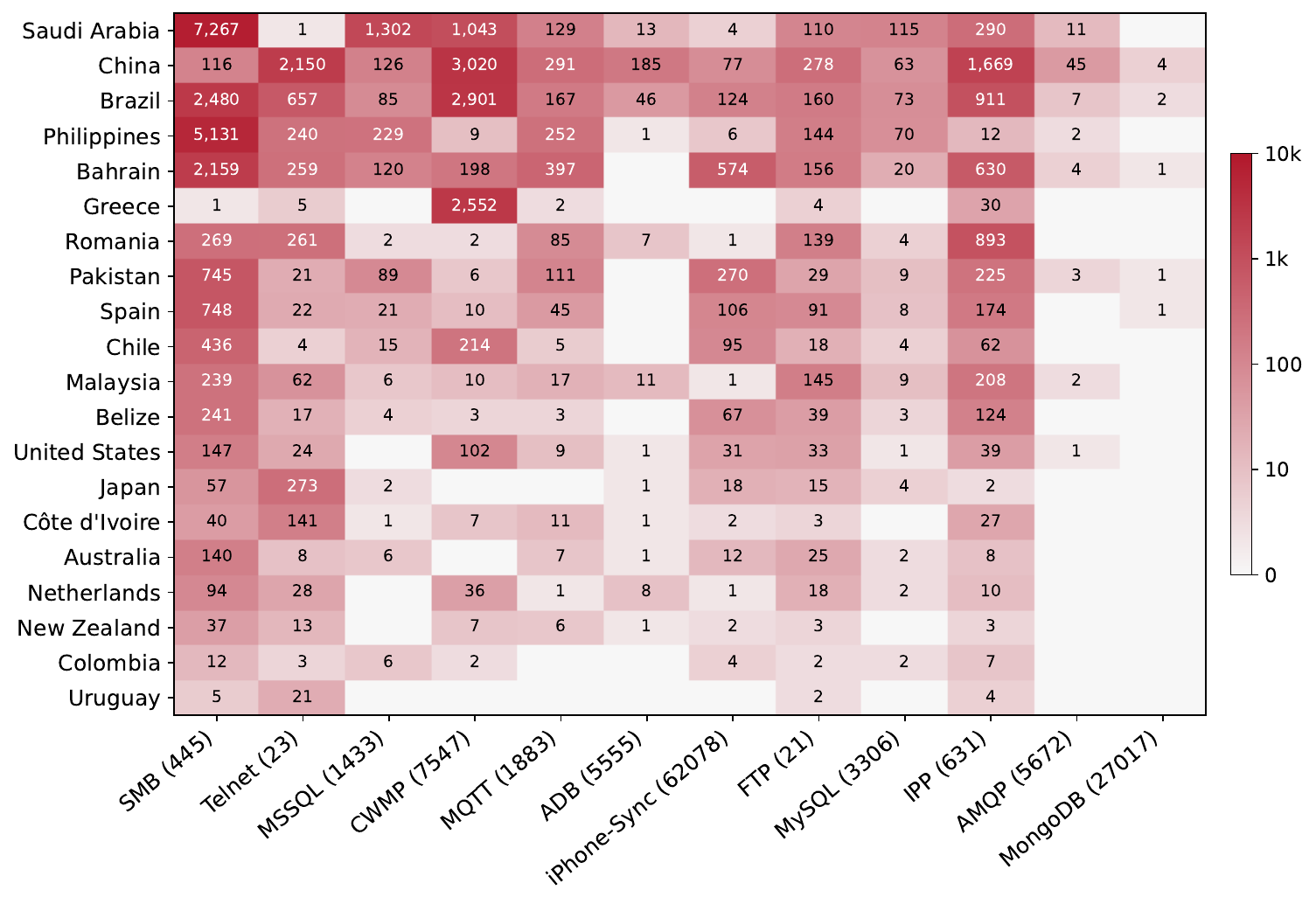}
   \Description{Heatmap of countries against protocols, where each cell counts unique internal IPv6 addresses responding to a protocol that their paired external address does not.}
    \caption{Unique internal IPv6 addresses that respond to a given protocol
    when their paired external address from the same network prefix does not.}
  \label{fig:geo-heatmap}
\end{figure}

\begin{table}[tb]
\caption{SSH server versions on internal-only IPv6 addresses (internal responds to SSH;
  paired external does not), with known CVEs.
  Versions with $<100$ addresses omitted.
  CVSS is v3 where scored, otherwise v2.
  CVE references: \textbf{1}~CVE-2013-4421: DoS via memory exhaustion, CVSS~5.0 (v2); \textbf{2}~CVE-2013-4434: user enumeration via timing, CVSS~5.0 (v2); \textbf{3}~CVE-2016-7406: format string RCE via username or host, CVSS~9.8; \textbf{4}~CVE-2016-7407: code execution via crafted key file in \texttt{dropbearconvert} (local tool), CVSS~9.8; \textbf{5}~CVE-2016-7408: code execution via crafted \texttt{-m}/\texttt{-c} argument in \texttt{dbclient} (client), CVSS~8.8; \textbf{6}~CVE-2016-7409: local memory disclosure via \texttt{-v} in DEBUG\_TRACE builds, CVSS~5.5; \textbf{7}~CVE-2018-15599: user enumeration, CVSS~5.3; \textbf{8}~CVE-2023-48795: Terrapin prefix truncation attack, CVSS~5.9; \textbf{9}~CVE-2024-6387: regreSSHion pre-auth RCE via SIGALRM race (glibc), CVSS~8.1.
}
\label{tab:ssh-versions}
\centering
\small
\setlength{\tabcolsep}{4pt}
\begin{tabular}{llrll}
\toprule
\textbf{Software} & \textbf{Version} & \textbf{IPs} & \textbf{Severity} & \textbf{CVEs} \\
\midrule
  Dropbear & \texttt{dropbear\_2012.55} & 8,417 & \textbf{Critical} & 1--7 \\
   & \texttt{dropbear\_2015.67} & 464 & \textbf{Critical} & 3--7 \\
   & \texttt{dropbear\_2014.63} & 373 & \textbf{Critical} & 3--7 \\
  OpenSSH & \texttt{OpenSSH\_9.2p1 (Debian)} & 550 & \textbf{High} & 8--9 \\
   & \texttt{OpenSSH\_8.9} & 308 & \textbf{High} & 8--9 \\
   & \texttt{OpenSSH\_8.9p1 (Ubuntu)} & 172 & \textbf{High} & 8--9 \\
   & \texttt{OpenSSH\_8.5} & 149 & \textbf{High} & 8--9 \\
   & \texttt{OpenSSH\_9.6p1 (Ubuntu)} & 223 & \textbf{High} & 9 \\
   & \texttt{OpenSSH\_9.6} & 115 & \textbf{High} & 9 \\
\bottomrule
\end{tabular}
\end{table}

\subsection{Internal-Only Reachability}

Finally, we look at networks in which both an internal and external address are
responsive, and where the internal addresses have unique
protocol reachability. That is, networks in which an internal address responds to
ports that the external address \emph{does not}.

Table~\ref{tab:internal-exposure} lists the protocols that are exposed only on
the internal \vsix address of a /56 but not the external \vsix address; this
scenario indicates that an attacker has increased access to in-home residential
devices over \vsix. HTTP/80 is the most common port/protocol combination, with
over 166k instances in which it is exposed on the internal address but not
external. The file sharing protocol SMB is exposed in over 21k instances on the
internal address but not the external address, potentially exposing private
information were an attacker to connect to the exposed shared directories. Other
protocols, such as SSH, may reveal information about the type of device or
permit fingerprinting or tracking from the server reply.

Table~\ref{tab:ssh-versions} lists a number of SSH software versions we received
from internal-only exposed SSH servers. We filtered these software versions for
only those with \emph{known high and critical CVEs}. Cumulatively, we find more
than 10k devices running exploitable versions of SSH within residential
networks, which make prime targets for botnets looking to move into
residential networks over \vsix.

Finally, Figure~\ref{fig:geo-heatmap} displays the number of internal-only
reachable port/protocol combinations grouped by the WHOIS country the internal
address is from. In Saudi Arabia, the country with the largest number of
internal-only reachable ports, over 7,000 unique \vsix addresses respond to SMB,
as well as more than 1,000 to MSSQL and CWMP. In Greek residential networks we
probe, by contrast, over 2,500 unique internal addresses have CWMP open when the
external address does not.

%% file: casestudies.tex
\section{Case Studies}
\label{sec:casestudies}

\begin{table}[t]
    \caption{HP printers discovered in \vsix (our work) vs.\ all of \vfour
    (Shodan), by AS.}
    \label{tab:printers}
    \centering
    \small
    \setlength{\tabcolsep}{4pt}
    \begin{tabular}{rrlc}
    \toprule
    \multicolumn{1}{c}{\textbf{Count}} & \multicolumn{1}{c}{\textbf{\%}} & \textbf{AS} & \textbf{CC} \\
    \midrule
    \multicolumn{4}{@{}l}{\emph{IPv6 (this work)}} \\
     7,465 & 16.5 & STC (AS25019)             & SA \\
     6,165 & 13.7 & V.tal (AS7738)            & BR \\
     6,022 & 13.3 & V.tal (AS8167)            & BR \\
     5,878 & 13.0 & RCS \& RDS (AS8708)       & RO \\
     2,274 &  5.0 & Chinanet (AS4134)         & CN \\
    17,325 & 38.4 & 637 others                &    \\
    45,129 &  100 & \textbf{Total}            &    \\
    \midrule
    \multicolumn{4}{@{}l}{\emph{Shodan (all IPv4)}} \\
     1,147 & 27.8 & Korea Telecom (AS4766)    & KR \\
       300 &  7.3 & AT\&T (AS7018)            & US \\
       124 &  3.0 & SK Broadband (AS9318)     & KR \\
        83 &  2.0 & Orange (AS3215)           & FR \\
        66 &  1.6 & Deutsche Telekom (AS3320) & DE \\
     2,410 & 58.4 & 932 others                &    \\
     4,130 &  100 & \textbf{Total}            &    \\
    \bottomrule
    \end{tabular}
\end{table}

\begin{figure*}[t]
  \centering
    \begin{subfigure}[t]{0.45\linewidth}
    \includegraphics[width=\textwidth]{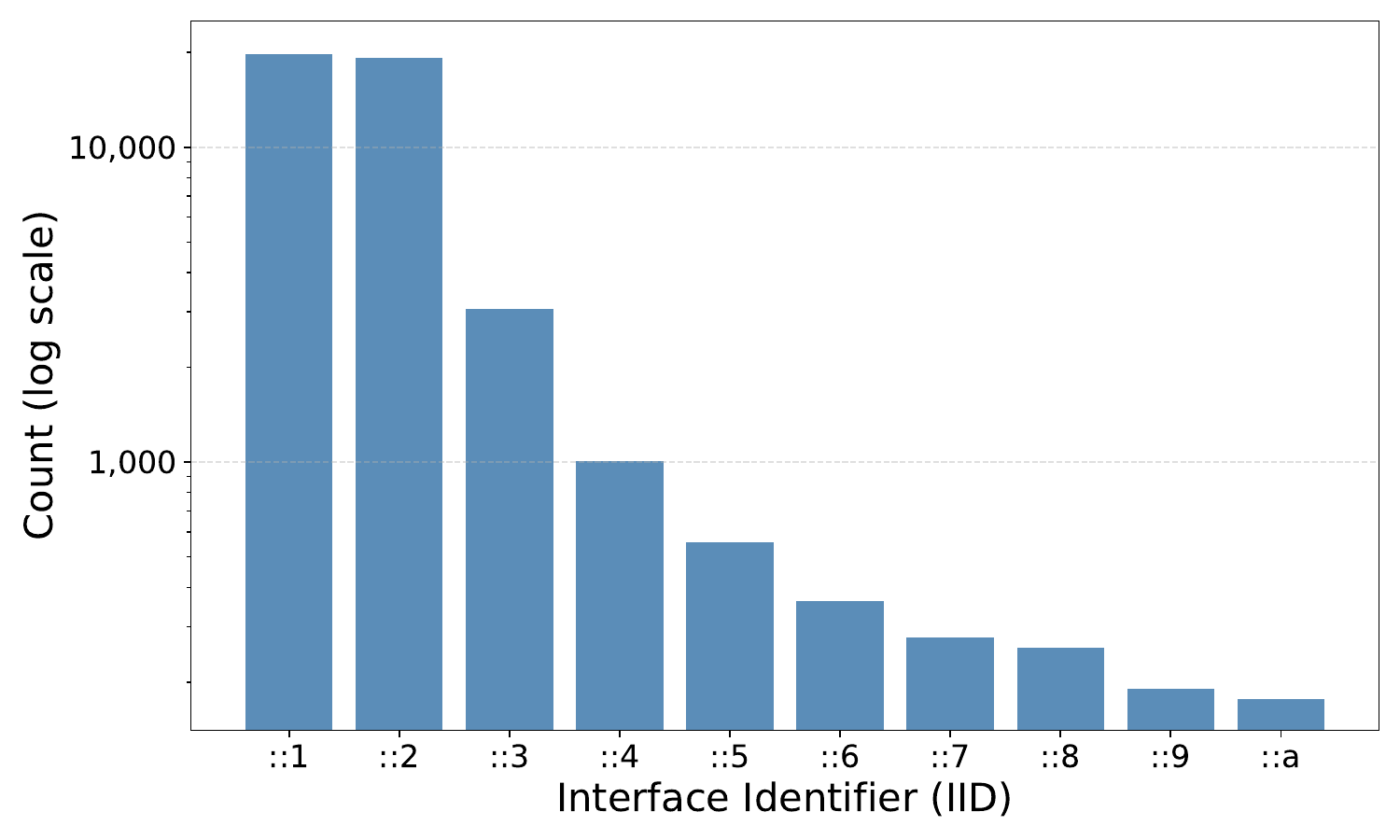}
    \Description{Bar chart with a log-scale y-axis of interface identifiers used by HP printers; ::2 is the most common, unlike the overall internal population where ::1 dominates.}
        \caption{Distribution of HP printer IIDs ($y$-axis logscale).}
      \label{fig:hp-printer-iid-bars}
  \end{subfigure}
  \hfill
    \begin{subfigure}[t]{0.45\linewidth}
        \includegraphics[width=\textwidth]{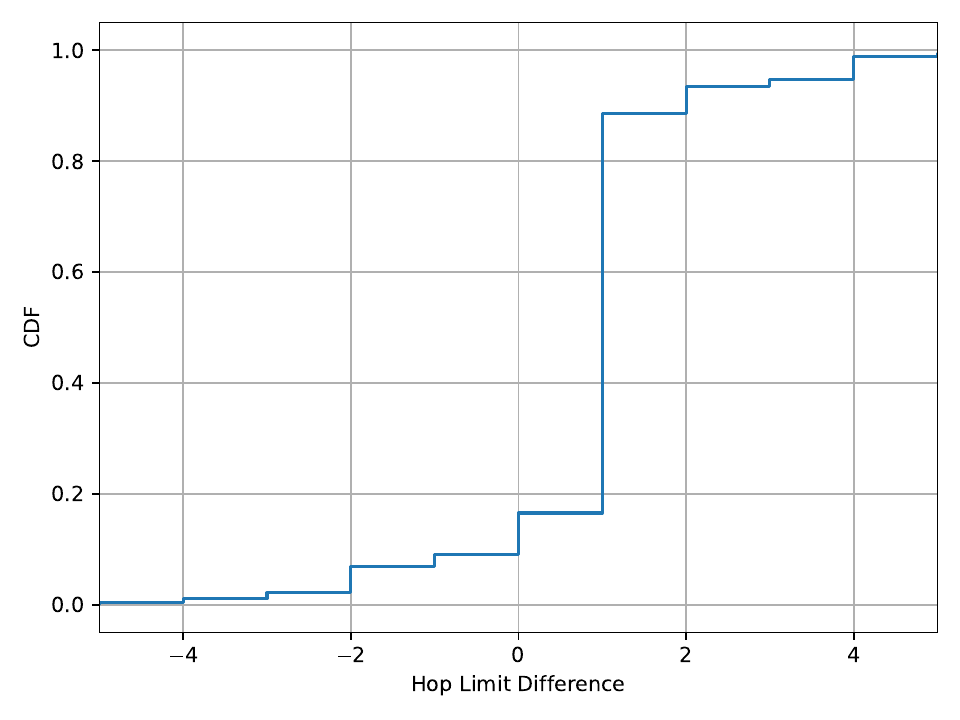}
        \Description{CDF of hop-limit distance from HP printers to the paired external address of their /56; roughly 72 percent are at distance plus one.}
        \caption{Hop-limit distance of HP printers to their associated external
        address.}
      \label{fig:hp-printer-hl-cdf}
  \end{subfigure}
        \caption{HP printer distributions of \vsix IIDs and hop-limit distance to
        their gateway differ significantly from the overall population.}
  \label{fig:printer-plots}
\end{figure*}

In this section,
we examine in more detail case studies of
devices that are definitively reachable on residential LANs.
These case studies highlight the current operational vulnerability that CPE routers
without \vsix stateful firewalling present to botnets.
In particular, we focus on printers, IP cameras, smart lighting systems, and Apple devices, as we can
unambiguously identify them via their unique responses as non-CPE.

\subsection{HP Printers}
\label{sec:printers}

Hewlett-Packard (HP) printers commonly run an embedded web server in order to
provide management and diagnostic capabilities.  We take advantage
of the fact that the HTTP responses from HP printers contain unique
signatures.
In general, HP printers include a \texttt{Server}
header that follows the format ``HP HTTP Server; $<$model name$>$; Serial Number:
$<$serial number$>$; Built: $<$build date$>$ {$<$build number$>$}''.  By filtering
for such responses within our HTTP ports 80 and 8080 and HTTPS 443 \zgrab results,
we can both
identify devices known to be in residential LANs and
correlate unique devices across protocols and time via the serial
number.

From 48,952 responsive printer addresses, we discover 45,129 unique devices
after removing 3,823 duplicate serial numbers that occur at multiple \vsix
addresses. Some devices are discoverable at multiple addresses on
the same LAN, such as the \texttt{::1} and \texttt{::4} addresses; we speculate
that these devices may have both a wired and \wifi connection. By contrast,
Williams~\etal discovered 351 HP printers (99\% fewer) using their TGA
technique~\cite{sixsense}.

\parhead{Verifying printers are inside residential LANs} Three pieces of evidence suggest that
these printers are indeed inside residential networks, as we designed our
methodology to discover. First, when both an HP printer (internal to the home LAN) and an external
address are responsive for the same probed network, the distribution of their \vsix IIDs in
Figure~\ref{fig:hp-printer-iid-bars} differs
substantially from the overall internal address IID distribution
(Figure~\ref{fig:internal-iid-bar} in the Appendix). Whereas the overall
internal IID distribution is overwhelmingly concentrated at \texttt{::1}
(94\% of internal addresses), for HP printers, 19,337 (43\%) have the
IID \texttt{::2}, and 25,255 (56\%) have an IID that is not \texttt{::1}.

Second, we examine how far, in hop-limit distance, HP printers sit from the
external \vsix address of the /56 we discover them in.
Figure~\ref{fig:hp-printer-hl-cdf} shows that, of the 48,127 printer addresses
with a responsive external address, nearly three-quarters (34,640, 72\%) are at a
hop-limit distance of +1 behind it. Contrast this with Figure~\ref{fig:ttls}, which shows the
distance behind the external address for all \icmpvsix-responsive internal
addresses. Of all internal/external address pairs, the internal address is at a distance of 0
from the external address 65\% of the time, while at a distance of +1 only in
17\% of cases.

Third, the most common models we discover are low-cost desk printers, such as
DeskJet series models; Table~\ref{tab:printer-models} in the Appendix lists the
most common.

\parhead{Comparison with \vfour}

To compare our discovered HP printers in \vsix with those discoverable in
\vfour, we searched Shodan for the same ``HP HTTP Server'' server banner that we
used to detect HP printers in \vsix. We start from 39,733 HP-printer banners
returned by Shodan and discard probable honeypots in four passes. First, we
drop any record Shodan itself tags as a honeypot. Second, we drop records whose
HTTP ``Server:'' header exceeds 500 characters, a common honeypot tell where the
response stitches together dozens of distinct server signatures (median header
length is 3,017 chars for tagged honeypots vs.\ 148 for legitimate printers).
Third, we drop records whose Server: header does not parse as the canonical HP
format ``HP HTTP Server; $<$model$>$; Serial Number: $<$serial$>$'', since real HP firmware
always emits this string. Fourth, we drop any record whose serial number is
observed in more than one country, since real serials are globally unique and
cross-country recurrence is the signature of a replayed honeypot banner---three
serials alone (0123456789, TH73P4S0Q0063T, CN81HC6060) account for over 16,000
records spanning 22--51 countries each. After these filters we deduplicate by
serial number, since one printer is often scanned on multiple HTTP ports (80,
443, 8080, 631), leaving 4,130 unique IPv4 HP printers.

The upper half of Table~\ref{tab:printers} lists the HP printers we discovered by
AS and country (as determined using Team Cymru's AS lookup~\cite{cymru2008ip}).
The lower half contrasts with scans of the entire \vfour space conducted by
Shodan~\cite{shodan}.
We find significantly more (11$\times$) total
printers than Shodan does in the entire \vfour space. Moreover, our discovery
technique appears complementary to exhaustive \vfour scans, as we find more HP
printers in several individual countries than exhaustive \vfour scanning does
globally. Further, only 3 distinct serial numbers are found in both datasets,
indicating that the overlap between the two scanning techniques is nearly
disjoint.

\begin{figure}[t]
 \centering
 \includegraphics[width=\linewidth]{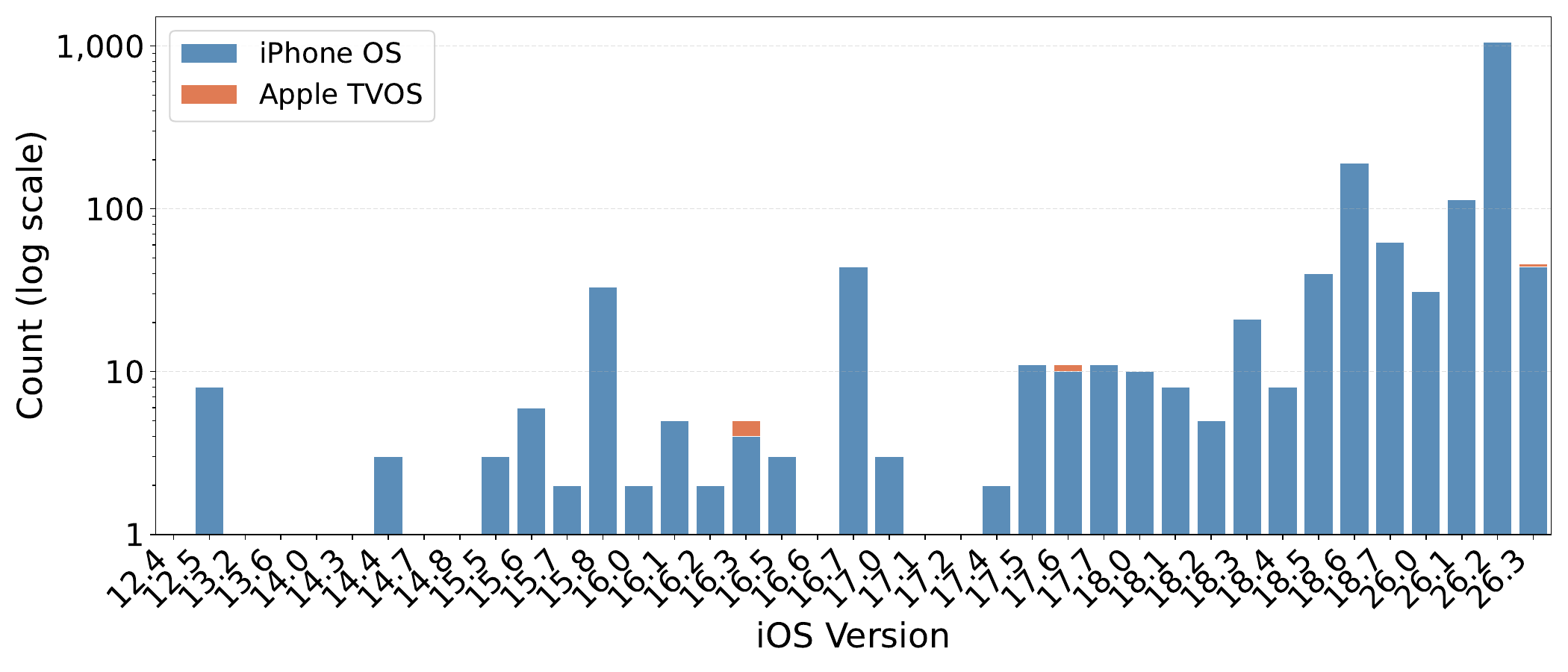}
 \Description{Histogram of iOS and tvOS versions reported by 1,749 Apple devices, with iOS 26.2 the most common and visible spikes at 15.8, 16.7 and 18.7.}
 \caption{iOS and tvOS versions of 1,749 Apple devices.}
 \label{fig:ios}
\end{figure}

\subsection{Apple Devices}

Next, we examine the prevalence of Apple devices within residential networks
using a custom \zgrab module we created. \texttt{lockdownd} is
an Apple service that runs on devices such as iPhones, iPads, and Apple
TVs~\cite{usbmuxd, usbmuxdapple} to transfer data over a USB cable; optionally,
users may opt in to enabling this service on TCP port 62078 to sync over
their local networks (\texttt{nmap} lists this TCP port as ``iphone-sync'', terminology we
borrow). While devices must be on the same iCloud account to transfer data, this
process will provide the device's operating system version without
authentication. We release this \zgrab module for other researchers
(Appendix~\ref{sec:open-science}).

We discovered 1,749 unique residential \vsix addresses running this service.
This is new exposure: iPhone-Sync is a LAN service that \vfour NAT makes
unreachable from the Internet unless a user deliberately forwards the port,
and an Internet-wide \vfour scan we conducted found only 850 such
hosts.
Figure~\ref{fig:ios} is a histogram of the iOS and tvOS versions we discovered
through our iPhone-Sync \zgrab scans. Notably, there are several versions of iOS
that are much more prevalent than others. iOS 26.2 was the latest release at the
time of these scans; this is reflected in its dominance over other release
versions.
Several dozen iPhones were running the beta iOS
version 26.3. Interestingly, we also note disproportionate numbers of iPhones
running iOS versions 15.8, 16.7, and 18.7. We attribute these disproportionate
version counts to the inability of certain iPhone models to be upgraded beyond
a fixed iOS version. For instance, the iPhone 6s, SE (1st generation), and 7
cannot be upgraded to iOS 16~\cite{iosversions}.

Like the HP printers in \S\ref{sec:printers}, Apple devices are more likely to
have IIDs other than \texttt{::1}. Some 324 (19\%) Apple devices we discovered
had the IID \texttt{::2} and 272 (16\%) had \texttt{::3}, while 319 (18\%)
had \texttt{::1}.

\subsection{IoT Devices}
\label{sec:iot}

We discovered a wide variety of IoT devices during our measurement campaign.

\parhead{Dahua IP camera systems} We discovered 4,659 internal \vsix addresses
for Dahua Technology~\cite{dahua} IP cameras by filtering for a fingerprint
unique to the HTML of these cameras' login pages (``Dahua'' appears only in an
image, so we filtered for the string \texttt{app-name=``cameraNewConfig''}). Like
HP printers, these devices were present in many countries but were
disproportionately located in a small number of them. Unlike HP printers, whose
top networks span the Middle East, Latin America, and Europe, the majority of
these devices were located in a single country.

Of the 4,659 Dahua camera systems we discovered, 3,196 (69\%) were located in
Brazil, according to Team Cymru's IP-to-ASN service. Another 913 (20\%) were
located in Romania. The remaining 550 (12\%) were located in another 15
countries. Notably, a search for the HTML fingerprint we used to identify these
devices returned no results on Shodan.

\parhead{Nanoleaf smart lights} Nanoleaf smart lighting systems~\cite{nanoleaf} are another clear example of
LAN-internal residential devices we discovered in our measurement campaign. In
our measurements, we discovered 802 Nanoleaf devices in 24 countries across 35
ASes, as determined by Team Cymru.
Like iPhone-Sync, this is a LAN service that \vfour NAT makes unreachable
unless deliberately forwarded.

These devices are recognizable due to their index page containing a link to
upload new firmware. While the device indicates that the user must physically
press a button on the lighting systems to effect the upgrade,
Internet-accessible IoT devices that accept uploads from arbitrary hosts
constitute a security risk that \vfour's NAT prevents. A majority of these
devices (443, 55\%) are found in Saudi Arabian networks, and 126 (16\%) are found in
Romania.

%% file: mitigations.tex
\section{Mitigations}
\label{sec:mitigations}

Mitigating the \vsix exposure of devices targeted by botnets requires the
assistance of multiple stakeholders.

\Parhead{Gateway vendors: Enable firewalls by default}
Although many residential gateway routers have firewalling
capabilities~\cite{olson2021natting}, our empirical results show that millions of gateway
routers do not turn them on.
A stateful firewall that drops unsolicited inbound traffic provides the same
protection that IPv4 NAT provided by accident.
To the extent that manufacturers have eschewed it, we hope this work
demonstrates that discovering devices inside their customers' networks is no
longer a hypothetical attack, but is possible even in \vsix's vast address
space.

\parhead{ISPs: Require CPE vendor firewalling}
Many residential gateways are supplied to customers by their ISP. If ISPs require
stateful \vsix firewalling of the vendors whose equipment they sell or lease,
this mitigation reaches users who never need to change a setting themselves.
ISPs could also drop certain types of unsolicited inbound traffic to residential
prefixes at their network boundaries. However, this mitigation has the potential
to cause collateral damage for users who knowingly wish to expose services to
the Internet.

\parhead{Users: Enable stateful firewalling}
We recommend that technical users check the status of their gateway's IPV6
firewall through its administrative web console, if it has one. If firewalling
options are exposed, users should ensure that they are enabled. While not all
CPE devices expose this level of control, those that do offer their owners a
mechanism to control their exposure to unsolicited inbound scanning.

\parhead{Devices: Reject low-entropy leases}
We were able to connect to many devices (e.g., iPhones) despite the fact that
they were following the best practices of generating random IIDs.
This was because they were \emph{also} accepting low-entropy IIDs, presumably
from their local DHCPv6 server.
A short-term mitigation for such devices would be to simply reject (or
significantly restrict inbound connections to) any low-entropy IIDs if they
already have a high-entropy IID.
This mitigation would need to be implemented by operating systems. There
is some precedent for doing so: Android does not support
DHCPv6 address assignment at all and instead relies on SLAAC~\cite{androiddhcpv6}.
We reiterate the suggestion of RFC~7707~\cite{rfc7707} for DHCPv6 servers to
assign random addresses drawn from a large pool.

\parhead{Operators: Monitor for IPv6 attacks}
Finally, operators should monitor IPv6 networks for attack traffic as they
monitor IPv4 today. However, based
on our experiences from actively scanning both IPv4 and IPv6 networks,
it appears that operators may not be monitoring them equally.
During our IPv4 scans, we received several opt-out requests and notifications
that our vantage point had been infected by malware. This was not an intended
goal of our measurements; on the contrary, we took steps to minimize how
malicious our traffic might look and the impact that it would have
on scanned networks or devices.
Yet the same protocol scans that produced alerts over IPv4---some within four
minutes of starting---produced not one over IPv6.
This raises a question for future work: are operators
monitoring their networks for malicious IPv6 traffic commensurately with how
they monitor IPv4?

%% file: conclusion.tex
\section{Conclusions}
\label{sec:conclusion}

We showed that IPv6 networks can be an easy and effective addition to IoT
botnets' arsenal.
Using a publicly available dataset of active /48s and a straightforward
scanning methodology that could run on virtually any device, we showed
that we can reach millions of services on residential networks.
We also demonstrated that, unlike with IPv4, we can reach \emph{inside} of the
networks---something that required sophisticated attacks or misconfigurations in IPv4 to
accomplish~\cite{slipstream}.

Our results paint a grim picture of a looming threat for the Internet.
At the time of writing, about half of the users reaching Google servers
do so over IPv6~\cite{googleipv6}. This change is---by design---largely
transparent to residential Internet subscribers.
In IPv4, NAT acted as an accidental but de facto firewall for residential
customers.
With NAT gone in IPv6, unwitting users must now rely on firewalls
and ephemeral, randomized IIDs to protect their networks
from being scanned.
Unfortunately, our results show that randomized IIDs are of limited utility, as
even devices that generate random IIDs (such as iPhones) still frequently obtain
DHCPv6 leases, which often give low-order IIDs (e.g., \texttt{::1} and
\texttt{::2}). And our results show that, in many cases, common CPE devices
permit unsolicited, inbound scan traffic into users' homes, where future
generations of botnets may find vulnerable devices ripe for exploitation.

To summarize our results simply: with their transition to IPv6, residential
networks are now more vulnerable to IoT botnet scanning.
We proposed new mitigations (and reiterated old ones) that users, service
providers, and device manufacturers can implement in order to better protect
their \vsix home networks.

%% file: appendix.tex
\section{Open Science}
\label{sec:open-science}

We support the goals of open and reproducible science. Toward that end, our artifacts are available at
\url{https://github.com/Network-Security-Privacy-Research/Where-Have-All-The-Firewalls-Gone-Artifacts}.
This includes the active /48s we classified as residential (\S\ref{subsec:residential});
our target-generation software for \zmapsix and our internal/external
classification code (\S\ref{subsec:scans}); and our custom iPhone-Sync and ADB
\zgrab modules.

\input{ethics}

\section{NTP Pool Servers}
\label{sec:ntp-pool-servers}

We ran NTP Pool servers in the following countries:
Australia, Brazil, Cyprus, Estonia, France,
Germany, Hong Kong, Hungary, India, Israel, Japan, Kazakhstan,
Norway, Poland, Russia, Serbia, Singapore, South Africa, South Korea,
Spain, Türkiye, Ukraine, United Arab Emirates, United Kingdom,
and the United States.

\raggedbottom

\begin{figure*}[t]
  \centering
  \begin{subfigure}[t]{0.45\linewidth}
    \includegraphics[width=\linewidth]{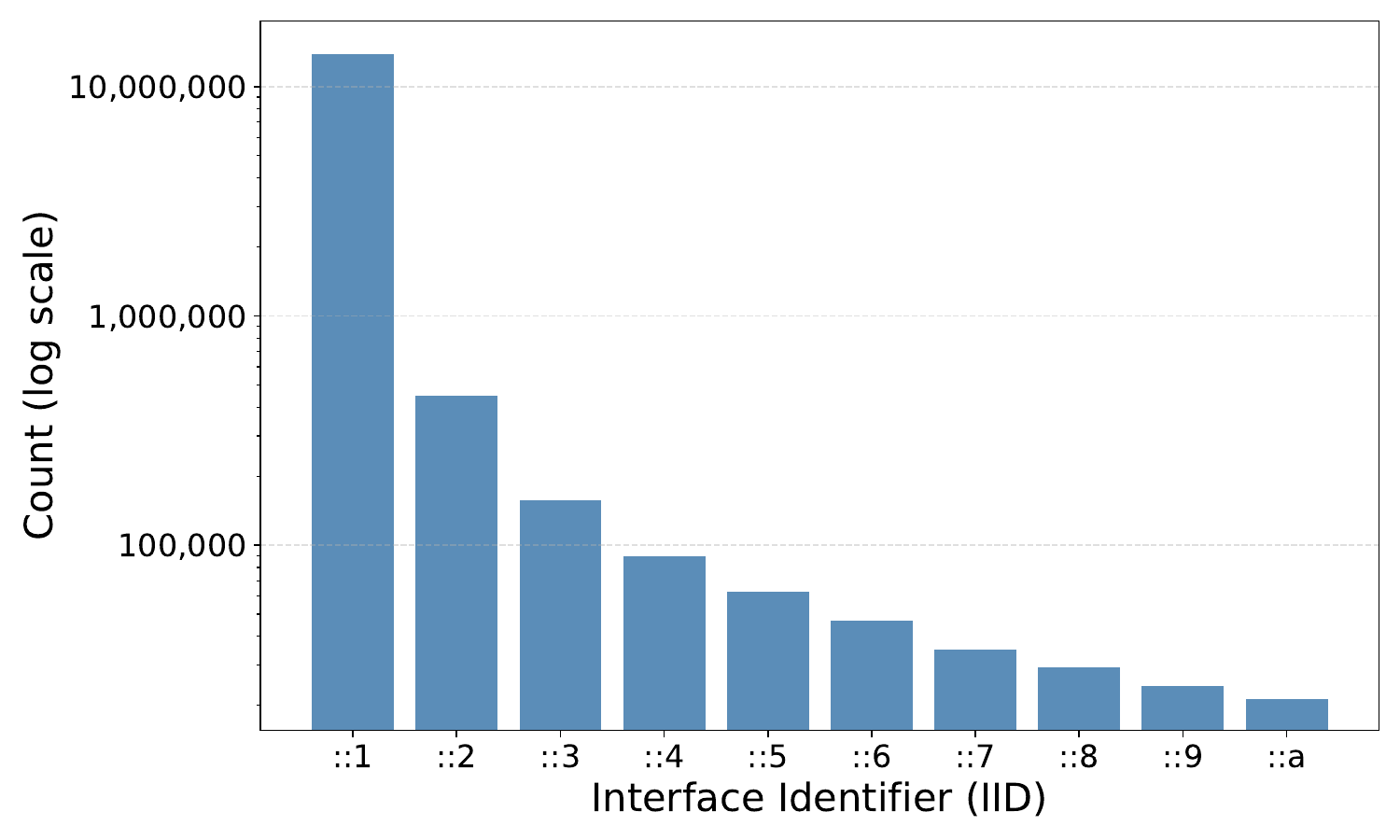}
    \Description{Bar chart with a log-scale y-axis of interface identifier counts among ICMPv6-responsive internal addresses; ::1 is about 31 times more common than ::2.}
    \caption{\icmpvsix-responsive internal hosts, \texttt{::1}--\texttt{::a}.}
    \label{fig:internal-iid-bar}
  \end{subfigure}
  \hfill
  \begin{subfigure}[t]{0.45\linewidth}
    \includegraphics[width=\linewidth]{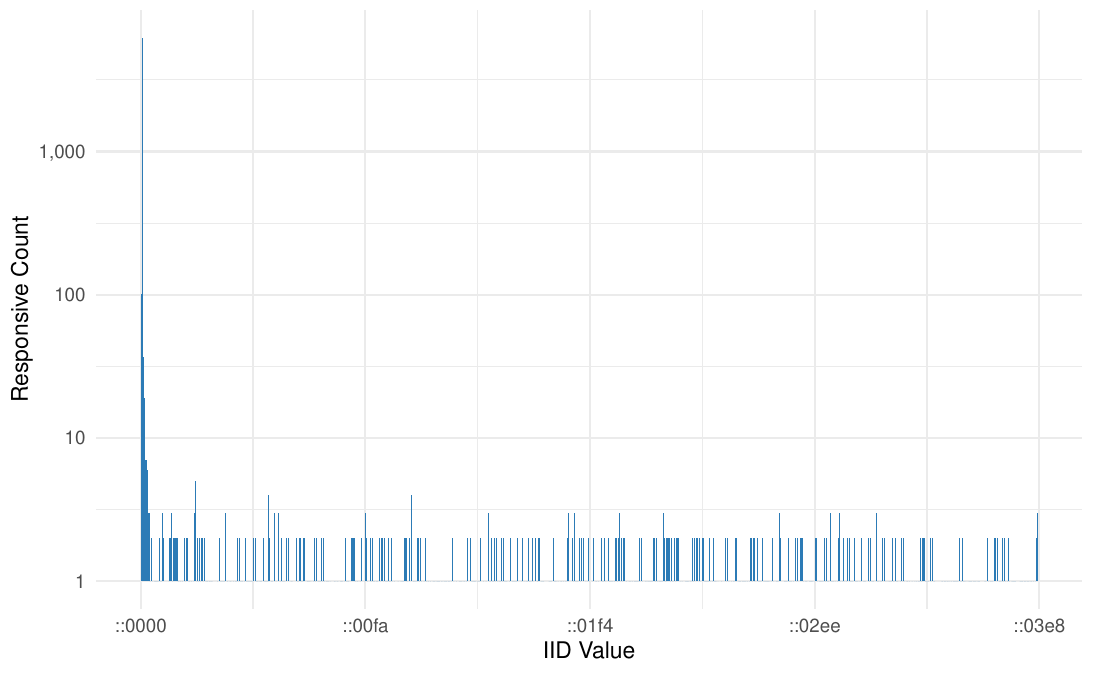}
    \Description{Bar chart with a log-scale y-axis of responsive addresses by interface identifier across the first 1,000 addresses of each /56 in 1,000 random /48s; 87.5 percent fall in the first ten.}
    \caption{First 1,000 IIDs of each /56 in 1,000 random /48s.}
    \label{fig:mini-experiment}
  \end{subfigure}
  \caption{Responsive addresses by IID ($y$-axes logscale).}
  \label{fig:iid-dists}
\end{figure*}

\newpage
\section{Additional Tables and Figures}
\label{sec:extra-tables}

Table~\ref{tab:services} provides a description of the port and protocol
combinations we scanned using \zgrab. Of note, some protocols (\eg TLS on port 8883, which we
probed as MQTTS) show up in some figures because the scanned device initiated
that protocol.

Table~\ref{tab:printer-models} shows the counts of the most common
printer models discovered on \vsix internal addresses. These counts
strongly indicate that we are indeed reaching devices internal to
residential networks.

  \begin{table}[H]
  \centering
      \caption{Services and ports we probed during our port scans (see
      \S\ref{subsec:scans}).}
  \label{tab:services}
  \setlength{\tabcolsep}{4pt}
      \begin{tabular}{@{}l p{1.1in} p{1.3in}@{}}
      \toprule
      \textbf{Protocol} & \textbf{Port(s)} & \textbf{Description} \\
      \midrule
      CWMP        & 7547                 & ISP management of CPE. \\
      HTTPS       & 443, 8443            & Primary and alternate port. \\
      HTTP        & 80, 5000, 8000, 8008, 8080, 8081 & Primary and alternate ports. \\
      TLS         & 443, 8443            & Bare TLS handshake. \\
      iPhone-Sync & 62078                & iOS sync protocol. \\
      IPP         & 631                  & Internet Printing Protocol. \\
      MQTT        & 1883                 & Message queue protocol. \\
      MQTTS       & 8883                 & MQTT over TLS. \\
      AMQP        & 5672                 & Message queue protocol. \\
      FTP         & 21                   & File transfer, command port. \\
      Telnet      & 23, 2323             & Remote access. \\
      SSH         & 22, 2222             & Secure shell. \\
      SMB         & 445                  & Windows file sharing. \\
      ADB         & 5555                 & Android Debug Bridge. \\
      MongoDB     & 27017                & MongoDB database. \\
      MSSQL       & 1433                 & Microsoft SQL database. \\
      MySQL       & 3306                 & MySQL database. \\
      NTP         & UDP/123              & Network Time Protocol. \\
      IMAP        & 143                  & Email retrieval. \\
      POP3        & 110                  & Email retrieval. \\
      SMTP        & 25                   & Mail transfer. \\
      \bottomrule
      \end{tabular}
  \end{table}

\begin{table}[H]
\caption{Top 5 HP printer models on internal \vsix addresses by count,
    deduplicated by serial number.}
\label{tab:printer-models}
\centering
\begin{tabular}{lrr}
\toprule
    \textbf{Model} & \textbf{Count} & \textbf{\%} \\
\midrule
HP DeskJet 2700 series - 7FR22A & 3,445 & 7.63 \\
HP Ink Tank Wireless 410 series - Z4B55A & 3,346 & 7.41 \\
HP Smart Tank 580-590 series - 4A8D5A & 3,330 & 7.38 \\
HP Smart Tank 580-590 series - 1F3Y2A & 3,083 & 6.83 \\
HP DeskJet 2800 series - 6W7G2A & 2,394 & 5.30 \\
Others & 29,531 & 65.44 \\
\midrule
    \textbf{Total} & \textbf{45,129} & \textbf{100.00} \\
\bottomrule
\end{tabular}
\end{table}

\begin{figure}[H]
    \centering
    \includegraphics[width=0.55\linewidth]{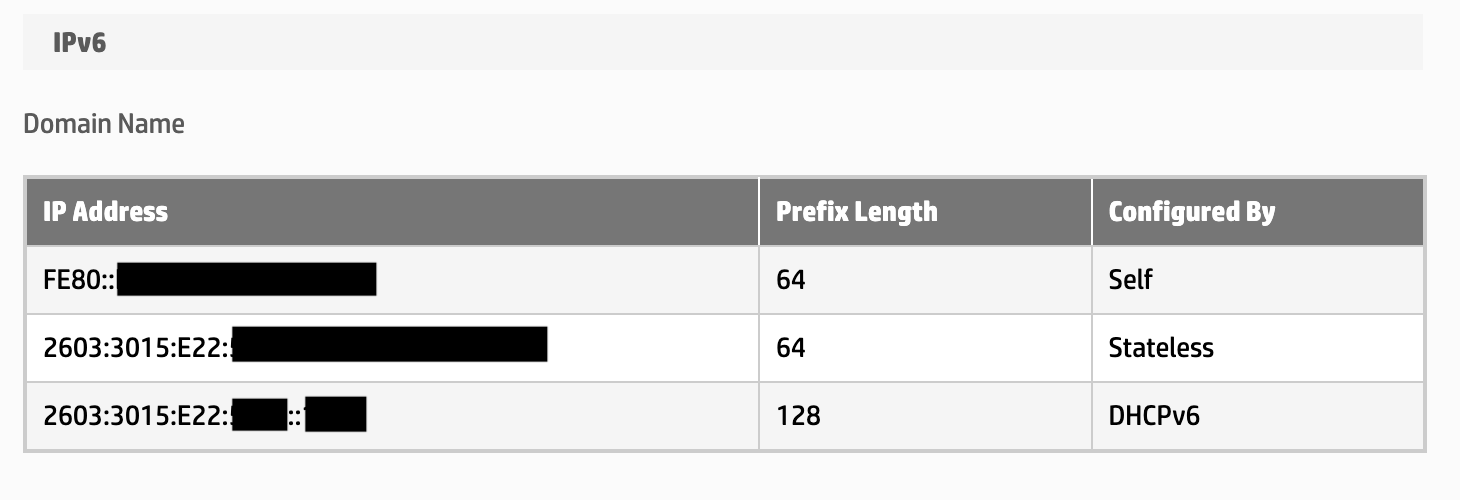}
    \Description{Screenshot of an HP printer's network status web page listing a DHCPv6 address, a random SLAAC address and a link-local address, with the addresses anonymized.}
    \caption{An HP printer status page. This device has both a DHCPv6-assigned
   address as well as a random, SLAAC-generated address. (Addresses
   anonymized for privacy.)}
    \label{fig:printerscreenshot}
\end{figure}

Figure~\ref{fig:internal-iid-bar} displays the distribution of internal \vsix
response IIDs. The IID \texttt{::1} is the most common, with 31 times more
responsive internal addresses than \texttt{::2}, the next-most common.

Figure~\ref{fig:mini-experiment} displays the number of responsive IIDs
observed from sending ICMPv6 Echo Requests to the first 1,000 addresses in each
/56 of 1,000 randomly chosen /48s. This motivated the choice in \S\ref{subsec:scans} to probe
only the first ten: 87.5\% of responsive addresses fall there.

Figure~\ref{fig:printerscreenshot} is an exemplar screenshot of the management
web page of a consumer-grade HP printer we identified over \vsix. This printer's
configuration page confirms that it has \emph{both} DHCPv6 and SLAAC \vsix
addresses, in addition to a link-local \vsix address. The DHCPv6 address, unlike
the random SLAAC address, is easily discoverable by probing for addresses with
only low IID bits set.

%% file: ethics.tex
\section{Ethical Considerations}
\label{sec:ethics}

Our institution's IRB determined this work exempt and not human-subjects
research.
We treat that determination as necessary but not sufficient, and summarize here,
following the Menlo Report's principles~\cite{menlo}, the considerations that
governed our data collection and handling.

\parhead{Basis for proceeding}
Our probes are benign, standard protocol handshakes of the kind Internet
measurement has relied on for over a decade~\cite{durumeric2013zmap}, sent to
addresses that are publicly routable by design of the ISPs that assign them.
We weighed the cost of a small number of packets to each network against the status
quo---millions of in-home devices reachable without their owners'
knowledge, remediable only once demonstrated---and judged that the
beneficence of identifying the vulnerability and providing concrete
remediation (\S\ref{sec:mitigations}) outweighed the potential harms.
We identified four potential harms: destabilizing or otherwise negatively
affecting a scanned device, causing excessive or undue load on a network,
collecting personal information, and collateral blocking of our provider's other
tenants.

\parhead{Minimizing harm}
We follow established best practices~\cite{durumeric2013zmap,sixsense}:
randomized targets, rate limiting, an explanatory web page on every scanning
host, and honored every opt-out request we received.
We port-scanned only addresses that first responded to ICMPv6, probed only ten
IIDs plus one alias check per /56, and dropped aliased networks outright
(\S\ref{subsec:scans}).
We scanned using RFC-compliant, well-known protocols unlikely to negatively
impact a device, and have disclosed our findings to a major router vendor, who
acknowledged that some of its products are affected; disclosure to other
manufacturers is ongoing.
Our scanning platform does not explicitly prohibit scanning; we retained
our IPv6 prefix for months after the campaign without scanning so that any block
placed on it would not fall on a later tenant.

\parhead{NTP Pool}
Clients that contact an NTP Pool server for time synchronization have not
consented to be port scanned.
As part of our stakeholder analysis we therefore consulted the Pool operators
before running our servers, and were told explicitly that scanning back the
/128s would violate the community's standards, but that /48s were acceptable
because they provide sufficient anonymity: residential customers are
typically allocated /56s, /60s, or /64s.
We retained only /48s, as Rye and Levin did~\cite{rye2023hitlists}, and
release only /48s.
Active /48s are also obtainable with no Pool server at all, from BGP or public
datasets, so our starting point models a realistic attacker. We note that
botnets or their operators need not respect this prohibition and could obtain
full /128s from NTP Pool servers directly.

\parhead{Data}
We make minimal ZGrab2 requests and do not attempt authentication.
We treat every protocol response as potentially sensitive. The raw responses we
receive are stored on a machine only the authors can access and will not be
released.

\parhead{Justice}
The burden of our scans is small and falls uniformly on every network in our
seed data. The benefit of this study accrues to the same population that bears
it---residential IPv6 users, through vendor disclosure and the mitigations
suggested in \S\ref{sec:mitigations}.